\documentclass{icmart} 
\usepackage[matrix,arrow]{xy}
\usepackage{epsfig}
\usepackage{graphicx}
\usepackage{bm}
\theoremstyle{plain}

\theoremstyle{remark}

\newcommand{\sst}{\scriptscriptstyle}

\newcommand{\ka}{\kappa}
\renewcommand{\1}{\one}
\renewcommand{\2}{\two}
\newcommand{\3}{\three}

\renewcommand{\theequation}{\thesection.\arabic{equation}}

\newcommand{\pa}{\partial}
\newcommand{\ot}{\otimes}
\newcommand{\frt}{{\mathfrak t}}
\newcommand{\ra}{\to}

\newcommand{\fsl}{{\mathfrak s}{\mathfrak l}}

\newcommand{\al}{\alpha}

\newcommand{\be}{\beta}
\newcommand{\ga}{\gamma}
\newcommand{\Ga}{\Gamma}
\newcommand{\de}{\delta}
\newcommand{\De}{\Delta}
\newcommand{\ep}{\epsilon}
\newcommand{\la}{\lambda}

\newcommand{\si}{\sigma}

\newcommand{\vf}{\varphi}

\newcommand{\CA}{{\mathcal A}}
\newcommand{\CB}{{\mathcal B}}

\newcommand{\CC}{{\mathcal C}}
\newcommand{\CD}{{\mathcal D}}
\newcommand{\CE}{{\mathcal E}}
\newcommand{\CF}{{\mathcal F}}

\newcommand{\CH}{{\mathcal H}}

\newcommand{\CL}{{\mathcal L}}
\newcommand{\CM}{{\mathcal M}}
\newcommand{\CN}{{\mathcal N}}
\newcommand{\CO}{{\mathcal O}}
\newcommand{\CP}{{\mathcal P}}

\newcommand{\CS}{{\mathcal S}}
\newcommand{\CT}{{\mathcal T}}

\newcommand{\CU}{{\mathcal U}}
\newcommand{\CV}{{\mathcal V}}
\newcommand{\CW}{{\mathcal W}}
\newcommand{\CX}{{\mathcal X}}
\newcommand{\CY}{{\mathcal Y}}
\newcommand{\CZ}{{\mathcal Z}}

\newcommand{\SB}{{\mathsf B}}

\newcommand{\SF}{{\mathsf F}}

\newcommand{\SH}{{\mathsf H}}

\newcommand{\sk}{{\mathsf k}}
\newcommand{\SL}{{\mathsf L}}

\newcommand{\SM}{{\mathsf M}}

\newcommand{\SR}{{\mathsf R}}

\newcommand{\ST}{{\mathsf T}}
\newcommand{\SU}{{\mathsf U}}

\newcommand{\SX}{{\mathsf X}}

\renewcommand{\sf}{{\mathsf f}}

\newcommand{\sll}{{\mathsf l}}

\newcommand{\BD}{{\mathbb D}}

\newcommand{\homsl}{\CM^{\BC}_{\rm char}(C)}
\newcommand{\homslr}{\CM^{\BR}_{\rm char}(C)}

\newcommand{\An}{{\mathfrak A}{\mathfrak n}}

\newcommand{\FV}{{\mathfrak V}}

\newcommand{\one}{{\mathfrak 1}}
\newcommand{\two}{{\mathfrak 2}}
\newcommand{\three}{{\mathfrak 3}}

\newcommand{\BF}{{\mathbb F}}

\newcommand{\BR}{{\mathbb R}}
\newcommand{\BH}{{\mathbb H}}

\newcommand{\BI}{{\mathbb I}}
\newcommand{\BC}{{\mathbb C}}

\newcommand{\BP}{{\mathbb P}}
\newcommand{\BS}{{\mathbb S}}

\newcommand{\BU}{{\mathbb U}}
\newcommand{\BZ}{{\mathbb Z}}

\newcommand{\rf}[1]{(\ref{#1})}

\newcommand{\Fus}[6]{F_{#5#6}^{}\big[\,{}^{#3}_{#4}\;{}^{#2}_{#1}\,\big]}

\newcommand{\nc}{\newcommand}

\nc{\rnc}{\renewcommand} \nc{\beq}{\begin{equation}}
\nc{\eeq}{\end{equation}} \nc{\beqa}{\begin{eqnarray}}
\nc{\eeqa}{\end{eqnarray}}

\title[Quantization of $\CM_{\rm flat}(C)$ and conformal field theory]{Quantization of moduli spaces of flat connections and Liouville theory}
\author[J. Teschner]{J. Teschner}
\contact[joerg.teschner@desy.de]{DESY Theory, Notkestr. 85, 22603 Hamburg, Germany}

\begin{document}
\begin{abstract}
{We review known results on the relations between 
conformal field theory,
the quantization of moduli spaces
of flat ${\rm PSL}(2,\BR)$-connections 
on Riemann surfaces, and the quantum Teichm\"uller theory.}\end{abstract}

\maketitle




\section{Introduction}
\setcounter{equation}{0}

The Teichm\"uller spaces $\CT(C)$ 
are the spaces of deformations of the 
complex structures on Riemann surfaces $C$. 
The classical uniformization theorem gives an alternative picture 
as spaces of constant negative curvature 
metrics modulo diffeomorphisms. Such metrics 
naturally define flat ${\rm PSL}(2,\BR)$-connections on $C$,
relating the Teichm\"uller spaces to the moduli spaces $\CM^{\BR}_{\rm flat}(C)$ 
of flat ${\rm PSL}(2,\BR)$-connections.

There are well-known connections between the 
Teichm\"uller spaces and the (complexified)
Lie-algebra of smooth vector fields on the unit circle. 
Cutting out a disc from a
Riemann surface $C$, and gluing it back after twisting by the flow
generated by a given vector field may generate changes 
of the complex structure of $C$. Both the spaces of functions on the Teichm\"uller spaces
and the dual to the space of vector fields on the unit 
circle have natural Poisson-structures which can be used
to formulate quantisation problems. Quantisation of the 
dual to the space of vector fields
on the unit circle gives the Virasoro algebra, the Lie algebra of symmetries
of any conformal field theory.
Despite the fact that
the existence of a relation between the quantisation of 
the Teichm\"uller spaces and conformal field theory may seem natural
from this point of view,
it has turned out to be nontrivial to establish such 
connections more precisely. The goal in this article will be
to outline what is currently known about the connections between
quantized moduli spaces of flat ${\rm PSL}(2,\BR)$-connections, 
quantum Teichm\"uller
theory, and conformal field theory\footnote{Here  
understood more precisely as representation theory of the Virasoro algebra 
with central charge $c>1$, corresponding to what is often called
Liouville theory in the physics literature.}.

The resulting picture appears to be of certain mathematical 
interest. It can in particular be seen as a first example for non-compact 
generalisations
of the known relations between conformal field theories, quantum 
groups and three-manifold invariants associated to compact (quantum-) groups. 
Indeed, many pieces of the resulting picture show deep analogies 
or even concrete relations to the harmonic analysis of non-compact groups.
Relations with three-dimensional hyperbolic geometry appear naturally,
coming from known relations between three-dimensional hyperbolic geometry and 
Teichm\"uller theory.
There are furthermore various connections with the theory
of classical and quantum
integrable models, including relations to the  isomonodromic
deformation problem. A unifying perspective was outlined in \cite{T10},
embedding such relations into a diamond of relations between
conformal field theory, the (classical and quantized) 
Hitchin moduli spaces, and the geometric Langlands correspondence.

There exist 
various applications of the mathematical results described
here in theoretical physics. They include relations to two-dimensional
quantum gravity and matrix models, (non-critical) string theory and various
relations to integrable models. Most strikingly, there even
exist relations to four-dimensional $\CN=2$-supersymmetric
gauge theories. Most direct seem to be the relations
to work of Alday, Gaiotto and Tachikawa \cite{AGT},
Gaiotto, Moore and Neitzke \cite{GMN2,GMN3},  
Nekrasov and Witten \cite{NW} and Nekrasov, Rosly and Shatashvili \cite{NRS}. 
These connections 
are described in \cite{TV2}.
 
This article concentrates on some
mathematical aspects of the connections between 
quantization of moduli spaces of flat connections and conformal field
theory.

\section{Moduli of flat ${\rm PSL}(2,\BR)$-connections and Teich\-m\"uller theory}
\label{sec:loops}

In this section we will briefly review the necessary background 
on the relevant moduli spaces $\CM_{\rm flat}(C)$ of flat connections and 
Teich\-m\"uller theory. The main goal will be to 
describe the algebra $\CA(C)\equiv{\rm Fun}^{\rm alg}(\CM_{\rm flat}(C))$
of algebraic functions on $\CM_{\rm flat}(C)$ in terms of 
generators and relations in a way that will be useful 
for the quantization.

We will consider Riemann surfaces $C=C_{g,n}$ of genus $g$ with $n$ 
marked points called punctures. In this article we will consider
connections having regular\footnote{The connection is gauge 
equivalent to a meromorphic connection with simple poles at the 
punctures.}  singularities at the punctures only.

\subsection{Flat connections and uniformization}

Let $\CM_{\rm flat}(C)$ be the moduli space of flat 
${\rm PSL}(2,\BC)$-connections modulo gauge transformations.
To each flat ${\rm PSL}(2,\BC)$-connection $\nabla=d+A$ we may associate
its holonomies $\rho(\ga)$ along closed curves $\ga$ 
as $\rho(\ga)=\CP\exp(\int_\ga A)$. The map
$\ga\mapsto \rho(\ga)$ defines a representation of the 
fundamental group $\pi_1(C)$ in 
${\rm PSL}(2,\BC)$, defining a point in the 
character variety 
\begin{equation}
\homsl:={\rm Hom}(\pi_1(C),{\rm PSL}(2,\BC))/{\rm PSL}(2,\BC)\,.
\end{equation}
The space $\homsl$ contains the real slice $\homslr$
which is known  to decompose into a finite set of 
connected components \cite{Hi,Go}. 

The uniformisation theorem allows us to represent any  Riemann
surface $C$ as a quotient of the upper half plane $\BU$ by certain
discrete subgroups $\Ga$ of ${\rm PSL}(2,\BR)$ called Fuchsian groups,
$C\simeq \BU/\Ga$. The Fuchsian subgroups $\Ga$ define representations
of $\pi_1(C)$ in ${\rm PSL}(2,\BR)$.
There is a distinguished connected component  
$\CM_{\rm char}^{\BR,0}(C)$ in $\CM_{\rm char}^{\BR}(C)$ containing 
all the Fuchsian groups $\Ga$ uniformising Riemann surfaces.  
This component corresponds to a connected 
component $\CM^\CT_{\rm flat}(C)$ in the moduli space $\CM^\BR_{\rm flat}(C)$ 
of flat ${\rm PSL}(2,\BR)$-connections on $C$. $\CM^\CT_{\rm flat}(C)$ is 
called Teichm\"uller component as
it is isomorphic to the Teichm\"uller space $\CT(C)$ \cite{Hi,Go}.

\subsection{Coordinates associated to triangulations}

There exist useful systems of coordinates for $\CM_{\rm flat}(C)$ associated to 
triangulations of $C$ if $C$ has at least one puncture. Coordinates 
of this type were introduced for  ${\CT}(C)$ in \cite{Pe}; 
the shear-coordinates introduced in \cite{F97} are closely related;
there exists a natural complexification  \cite{FG};
the following formulation is due to \cite{GMN2}.

Let $\tau$ be a triangulation of the surface $C$ such that
all vertices coincide with marked points on $C$. An edge
$e$ of $\tau$ separates two triangles defining a
quadrilateral $Q_e$ with corners being the marked points
$P_1,\ldots,P_4$. For a given connection $\nabla=d+A$,
let us choose four sections $s_i$, $i=1,2,3,4$ that are
horizontal in $Q_e$,
\begin{equation}
\nabla s_i = \left(d+A\right)s_i = 0\,.
\end{equation}
We shall furthermore assume that the sections $s_i$ are eigenvectors
of the monodromy around $P_i$. Out of
the sections $s_i$ form 
\begin{align}
\CX_e^{\tau} := -\frac{(s_1\wedge s_2)(s_3\wedge s_4)}{(s_2\wedge s_3)(s_4\wedge s_1)},
\end{align}
where all $s_i$, $i=1,2,3,4$  are evaluated at the same point $P\in
Q_e$. The ratio $\CX_e^{\tau}$ does not
depend on the choice of $P$.

There is a natural Poisson structure on $\CM_{\rm flat}(C)$
induced by the symplectic form $\Omega_{\rm AB}$ 
introduced by Atiyah and Bott. 
The Poisson bracket of 
the coordinates $\CX_e$ becomes very simple,
\begin{equation}\label{poisson}
\{\CX_e^{\tau}, \CX_{e'}^{\tau}\}
\,=\, n_{e,e'}\, \CX_{e'}^{\tau}\,{\CX_e^{\tau}}\,;
\end{equation}
the definition of $n_{e,e'}\in\{-2,-1,0,1,2\}$ is 
best described in terms of the fat graph $\hat{\frt}$ 
dual to the given triangulation 
$\frt$: It is the total intersection index of the edges
$\hat{e}$  and $\hat{e}'$ dual to $e$ and $e'$, respectively. 

There furthermore exists a simple description of  
the relations between  the coordinates
associated to different triangulations. If  
triangulation $\tau_e$ 
is obtained from $\tau$ by changing only the diagonal in the quadrangle
containing $e$, we have 
\begin{equation}\label{cluster}
\CX_{e'}^{\tau_e}\,=\,\left\{
\begin{aligned} &\CX_{e'}^{\tau}\,
\big(1+
(\CX_e^{\tau})^{-{\rm sgn}(n_{e'e})}\big)^{-n_{e'e}} \;\;&{\rm if}\;\;e'\neq e\,,\\
&(\CX_e^{\tau})^{-1}\;\;&{\rm if}\;\;{e'= e}\,.
\end{aligned}\right.
\end{equation}
Poisson bracket \rf{poisson} and transformation law  \rf{cluster}
reflect the cluster algebra structure 
that $\CM_{\rm flat}^{}(C)$ 
has \cite{FG}.

\subsection{Trace functions}

Useful coordinate functions for $\CM_{\rm flat}(C)$ are the trace functions
\begin{equation}\label{tracedef}
L_\ga:=\nu_\ga{\rm tr}(\rho(\ga))\,;
\end{equation}
the signs $\nu_\ga\in\{+1,-1\}$ will be chosen in such a way
that the restriction of $L_\ga$ to $\CM_{\rm flat}^{\BR}(C)$
is positive and larger than two.
It is possible
to show that the length $l_\ga$ of the geodesic on $\BH/\Ga$ 
isotopic to $\ga$ is related
to $L_{\ga}$ as
$L_{\ga}\,=\,2\cosh(l_\ga/2)$\,.
If the curves $\ga_r$ encircle the punctures $P_r$ on
$C=C_{g,n}$ for $r=1,\dots,n$, we will identify the 
surface $C$ with the surface with constant negative curvature metric obtained by cutting out $n$ discs 
having the geodesics isotopic to $\ga_r$ as boundaries.
 
There exists a natural complex structure on $\CM_{\rm flat}(C)$ 
which is such that the trace functions $L_{\ga}$ defined above
are {\it complex analytic}.

\subsubsection{Skein algebra}

Let $\CA(C)\simeq {\rm Fun}^{\rm alg}(\CM_{\rm flat}(C))$ be the  
commutative algebra  
of functions  on $\CM_{\rm flat}(C)$ 
generated by the coordinate functions $L_\ga$. We will explain how to
describe $\CA(C)$ in terms of generators and relations 

\begin{figure}[t]
\epsfxsize8cm
\centerline{\epsfbox{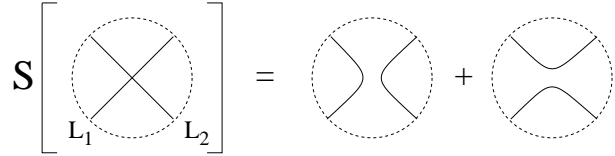}}
\caption{The symmetric 
smoothing operation}\label{skeinfig}\vspace{.3cm}
\end{figure}

The well-known relation ${\rm tr}(g){\rm tr}(h)={\rm tr}(gh)+{\rm tr}(gh^{-1})$
valid for any pair of $SL(2)$-matrices $g,h$ implies that the
trace functions satisfy the 
skein relations,
\begin{equation}\label{skeinrel}
L_{\ga_1} L_{\ga_2}\,=\,L_{S(\ga_1,\ga_2)}\,,
\end{equation}
where $S(\ga_1,\ga_2)$ is the curve obtained from 
$\ga_1$, $\ga_2$ by means of the smoothing operation,
defined as follows. The application of $S$ to a single intersection
point of $\ga_1$, $\ga_2$ is depicted in 
Figure \ref{skeinfig}.
The general result is obtained by
applying this rule at each intersection point, and summing the
results.

\subsubsection{Topological classification of closed curves}\label{Dehn}

A Riemann surface $C$ of genus $g$ with $n$ punctures may be cut into 
pairs of pants by cutting along $h:=3g-3+n$ non-intersecting simple 
closed curves $\ga=\{\ga_1,\dots,\ga_{h}\}$ on $C$. It will be
useful to supplement the collection of curves $\ga$ specifying a pants decomposition
by a three-valent graph $\Ga$ on $C$ which has exactly one vertex 
inside each pair of pants, and the three edges emanating 
from a given vertex each intersect exactly one of the boundaries of the pair of pants.
The pair of data $\si=(\ga,\Ga)$ will be called a pants
decomposition\footnote{The graph $\Ga$ allows us to distinguish pants
decompositions related by Dehn-twists, the operation to cut open along a curve 
$\ga_e\in\ga$, twisting by $2\pi$, and gluing back.}.

With the help of pants decompositions one may conveniently classify all
non-selfintersecting closed curves on $C$ up to homotopy \cite{De}. 
Recall that there is a unique curve $\ga_e\in\ga$ that intersects 
a given edge $e$ on $\Ga$ exactly once, and which does not intersect
any other edge. To a curve $\ga_e\in\ga$
let us associate the integers $(r_e,s_e)$ 
defined as follows.The integer $r_e$ is defined as 
the number of intersections between $\ga$ and 
the curve $\ga_e$. Having chosen an orientation 
for the edge $e$ we will define
$s_e$ to be the intersection index between $e$ and $\ga$. 

Dehn's theorem \cite{De}
ensures that the curve $\ga$ is up to homotopy uniquely classified by the
collection of integers $(r,s):e\mapsto (r_e,s_e)$, subject to the restrictions 
\begin{equation}
\begin{aligned}
{\rm (i)} \quad & 
r_e\geq 0\,,\\ {\rm (ii)} \quad & {\rm if}\;\;r_e=0\;\Rightarrow\;s_e\geq 0\,,\\
{\rm (iii)} \quad &
r_{e_\1}+r_{e_\2}+r_{e_\3}\in 2\BZ\;\,{\rm whenever}\;\,
\ga_{e_\1},\ga_{e_\2},\ga_{e_\3}\;\,\text{bound the same}\\
& \text{pair of pants}.
\end{aligned}
\end{equation}
We will use the notation $\ga_{(r,s)}$ for the geodesic which has
parameters $(r,s):e\mapsto (r_e,s_e)$.

\subsubsection{Generators} \label{sec:Gen}

As set of generators for $\CA(C)$ one may take the functions
$L_{(r,s)}\equiv L_{\ga_{(r,s)}}$. The skein relations
allow us to express arbitrary $L_{(r,s)}$ in terms
of a finite subset of the set of $L_{(r,s)}$. We shall now
describe convenient choices for sets of generators.

Let us note that 
to each internal\footnote{An internal edge does not end 
in a boundary component of $C$.} edge $e$ of the graph $\Ga$ of 
$\si=(\ga,\Ga)$
there corresponds a unique  curve $\ga_e$
in the cut system $\CC_\si$. There is a unique subsurface 
$C_{e}\hookrightarrow C$
isomorphic to either $C_{0,4}$ or $C_{1,1}$ that contains $\ga_e$ in 
the interior of $C_e$. The subsurface $C_e$ has boundary components
labeled by numbers $1,2,3,4$ 
if $C_e\simeq C_{0,4}$, and if $C_e\simeq C_{1,1}$
we will assign to the single boundary component the label $0$.

\begin{figure}[t]
\epsfxsize4cm
\centerline{\epsfbox{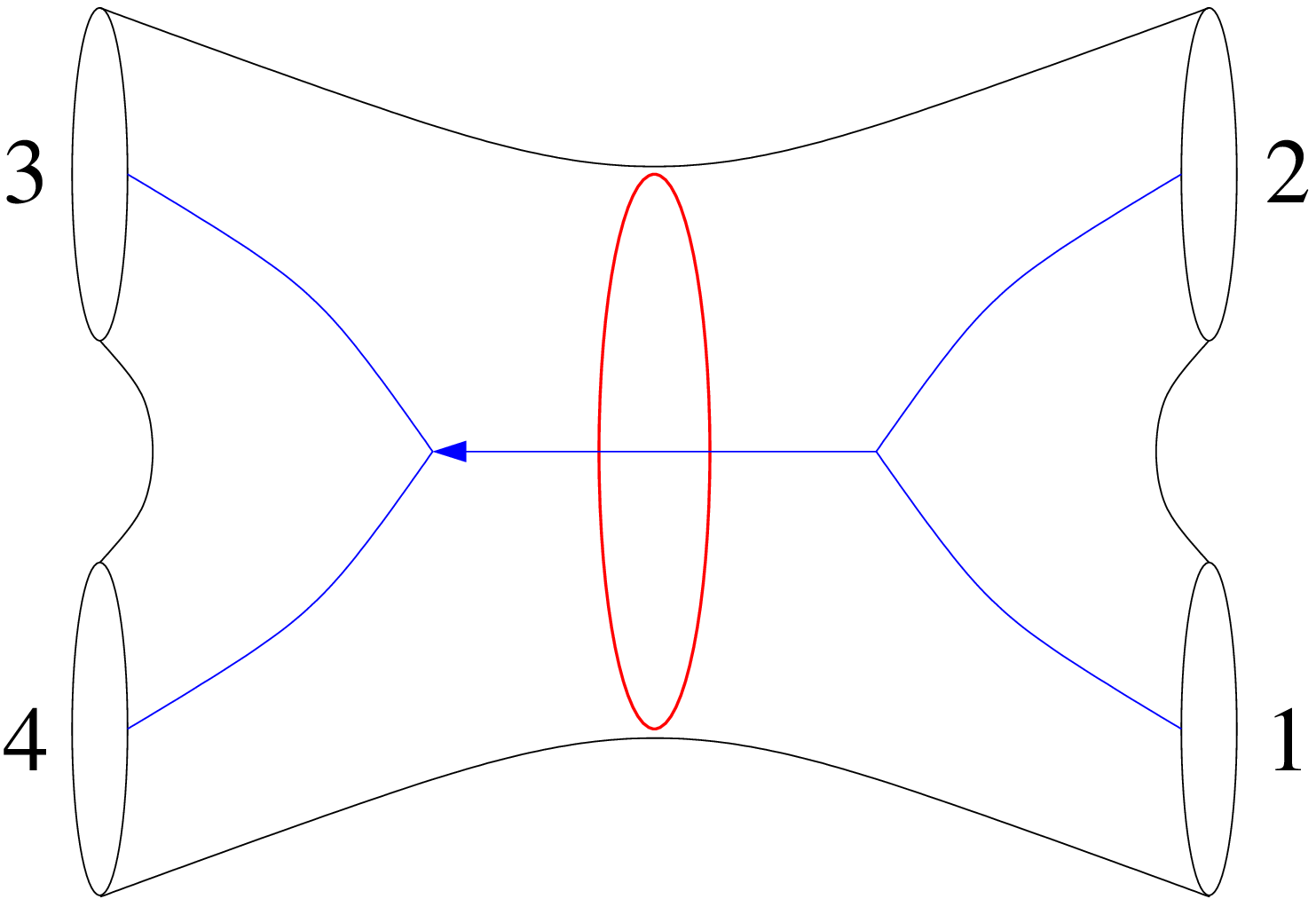}\hspace{.5cm}$\Longrightarrow$\hspace{.5cm}
\epsfxsize4cm\epsfbox{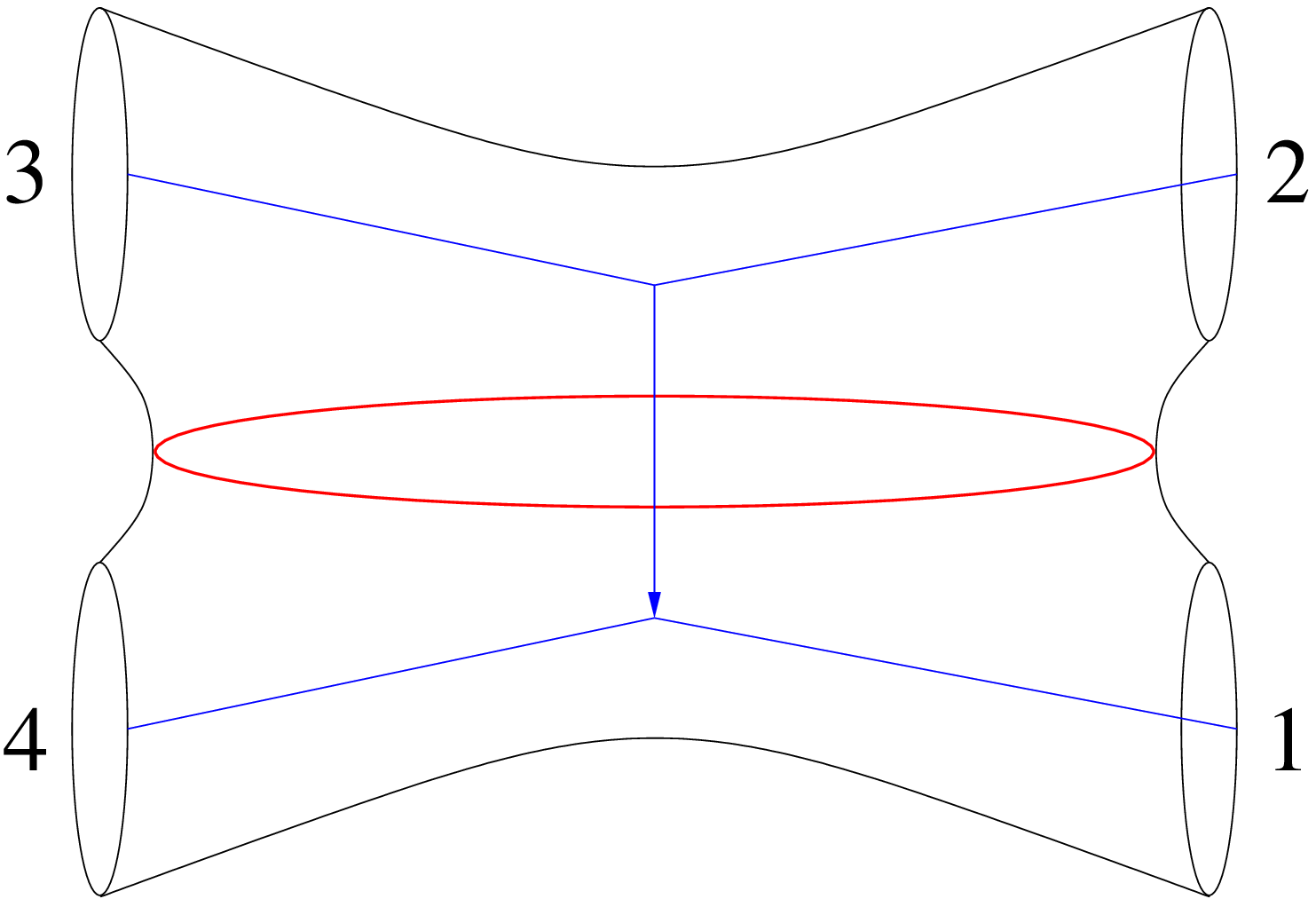}}
\caption{The geodesics $\ga^e_s$ and $\ga^e_t$ 
are the red curves on the left and right pieces of the figure. 
The change of pants decomposition from left to
right is called F-move}\label{fmove}\vspace{.3cm}
\end{figure}
\begin{figure}[t]
\epsfxsize5cm
\centerline{\epsfbox{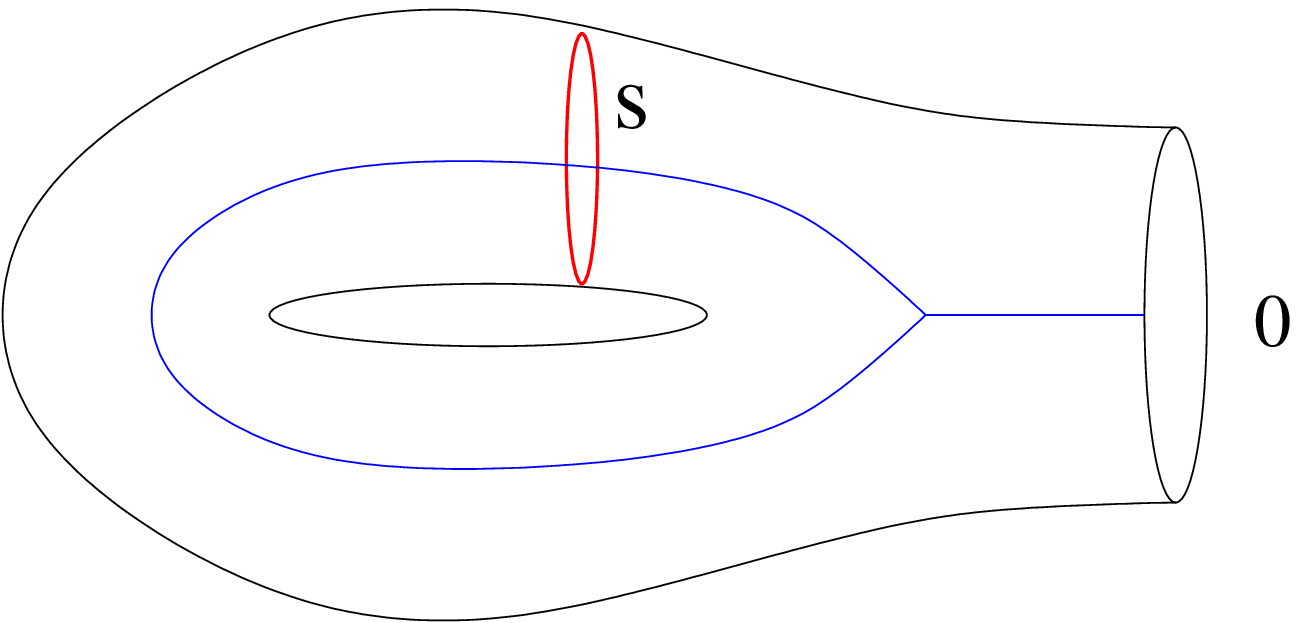}\hspace{.5cm}$\Longrightarrow$\hspace{.5cm}
\epsfxsize5cm\epsfbox{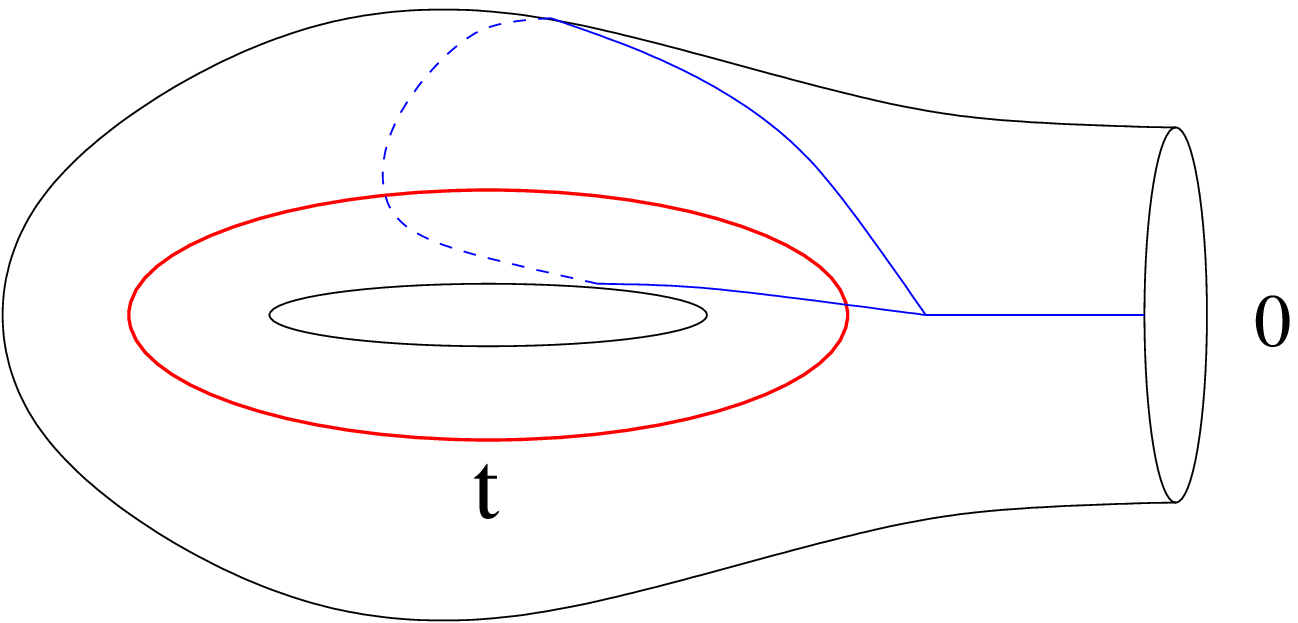}}
\caption{The geodesics $\ga^e_s$ and $\ga^e_t$ 
are the red curves on the left and right pieces of the figure. 
The change of pants decomposition from left to
right is called  S-move}\label{smove}\vspace{.3cm}
\end{figure}

For each edge $e$ let us introduce the geodesics $\ga^e_t$
which have Dehn parameters $(r^e,0)$, where
$r_{e'}^e=2\de_{e,e'}$ if $C_e\simeq C_{0,4}$ and 
$r_{e'}^e=\de_{e,e'}$ if $C_e\simeq C_{1,1}$. 
The geodesics $\ga^e_s$ and $\ga^e_t$
are depicted as red curves on the left and 
right halfs of Figures \ref{fmove}
and \ref{smove}, respectively. 
There furthermore exist unique
geodesics $\ga^e_u$ 
with Dehn parameters $(r^e,s^e)$, where $s_{e'}^e=\de_{e,e'}$.
We will denote the trace functions associates to $\ga_k^e$ by 
$L_{k}^e$, where $k\in\{s,t,u\}$.
The set
$\{L_s^e,L_t^e,L_u^e\,;\,\ga_e\in\ga\}$ generates $\CA(C)$.

\subsubsection{Relations}\label{sec:Rel}

These coordinates are not independent, though. Further relations
follow from the relations in $\pi_1(C)$.
It can be shown (see e.g. \cite{Go09} for a review) 
that any triple of
coordinate functions $L_s^e$, $L_t^e$ and $L_u^e$
satisfies an algebraic
relation of the form 
\begin{equation}
\label{algrel}
P_e(L_s^e,L_t^e,L_u^e)\,=\,0\,.
\end{equation}
The polynomial $P_e$ in \rf{algrel} is
for $C_e\simeq C_{0,4}$ explicitly given
as
\begin{align}\label{W04}
 P_e& (L_s, L_t, L_u) : = L_1L_2L_3L_4+L_s^2 + L_t^2 + L_u^2 +L_1^2+
L_2^2+L_3^2+L_4^2-4 \\
& \quad+L_s (L_3L_4 + L_1L_2) + L_t (L_2L_3 + L_1L_4) + L_u (L_1L_3 + L_2L_4)
-L_s L_t L_u \,,\nonumber\end{align}
while for $C_e\simeq C_{1,1}$ we take $P$ to be 
\begin{align}
 P_e(L_s, L_t, L_u) := L_s^2 + L_t^2 + L_u^2 - L_s L_t L_u + L_0-2\,.
\end{align}
In the expressions above we have denoted $L_{i}:=\nu_{\ga_i}
{\rm Tr}(\rho(\ga_i))$,
$i=0,1,2,3,4$, where $\ga_0$ is the geodesic representing the 
boundary of $C_{1,1}$, while
$\ga_i$, $i=1,2,3,4$ represent 
the boundary components of $C_{0,4}$. Generators $L^e_k$, $k\in\{s,t,u\}$, and relations \rf{algrel}
for all edges $e$ of $\Ga$ describe $\CM_{\rm flat}(C)$ as an algebraic variety.

\subsection{Poisson structure}

\begin{figure}[t]
\epsfxsize8cm
\centerline{\epsfbox{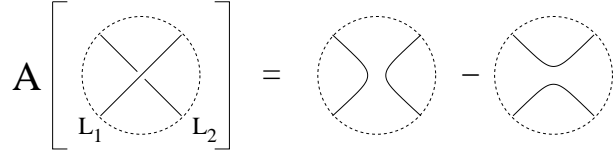}}
\caption{The anti-symmetric smoothing operation}\label{skeinfigas}\vspace{.3cm}
\end{figure}

There is also a natural Poisson bracket on $\CA(C)$ \cite{Go86}, 
defined such that 
\begin{equation}\label{goldbr}
\{\,L_{\ga_1}\,,\, L_{\ga_2}\,\}\,=\,L_{A(\ga_1,\ga_2)}\,,
\end{equation}
where $A(\ga_1,\ga_2)$ is the curve obtained from 
$\ga_1$, $\ga_2$ by means of the anti-symmetric smoothing operation,
defined as above, but replacing the rule depicted in 
Figure \ref{skeinfig} by the one depicted in Figure 
\ref{skeinfigas}. The Poisson structure \rf{goldbr} coincides
with the one induced 
from the symplectic form introduced by Atiyah and Bott.

The Poisson bracket $\{\,L_s^e\,,L_t^{e}\,\}$
can be written elegantly in the form \cite{NRS}
\begin{equation}\label{loopPB}
\{\,L_s^e\,,L_t^{e}\,\}\,=\,\frac{\pa}{\pa L_u^e} P_e(L_s^e,L_t^e,L_u^e)\,.
\end{equation}
It is remarkable that the same polynomial appears both in 
\rf{algrel} and in \rf{loopPB}, which indicates that the symplectic
structure on  $\CM_{\rm flat}(C)$ is compatible with its 
structure as algebraic variety.

It is sometimes useful to introduce Darboux-coordinates $e\mapsto (l_e,k_e)$ such that
$\{l_e,k_{e'}\}=\de_{ee'}$ and $L_s^e=2\cosh(l_e/2)$. The Fenchel-Nielsen coordinates 
for $\CT(C)$ are such coordinates. There is a natural complexification of the
Fenchel-Nielsen coordinates discussed in \cite{NRS}.

\section{Quantization
of $\CM_{\rm flat}^{\CT}(C)$}\label{q-Mflat}
\renewcommand{\CL}{{L}}

\setcounter{equation}{0}

We shall next review the quantization of the moduli spaces 
$\CM_{\rm flat}^{\CT}(C)$ that was constructed in \cite{T05,TV2}
based on the pioneering works \cite{F97,Ka1,CF}.

 \subsection{Quantization of coordinates associated to triangulations}

The simplicity of the Poisson brackets \rf{poisson}
of the coordinates $\CX_e^\frt$
makes part of the quantization quite simple. To each edge
$e$ of a triangulation $\frt$ of a Riemann surface $C_{g,n}$
associate the generator $\CX^{\frt}_e$ of a noncommutative algebra
$\CB_{\frt}$ which has generators $\CX^{\frt}_e$ and
relations
\begin{equation}\label{comm}
\CX^{\frt}_e,\CX^{\frt}_{e^\prime}=e^{2\pi ib^2 n_{ee'}} 
\CX^{\frt}_{e'}\,\CX^{\frt}_e\,,
\end{equation}
the integers $n_{ee'}$ coincide with the structure 
constants of the Poisson algebra \rf{poisson}, and we have introduced the
notation $b^2$ for the deformation parameter traditionally denoted
$\hbar$.

Note furthermore that the variables $\CX_e$ are positive
for the Teichm\"uller component. This motivates us to consider representations
$\pi_\frt$ of $\CB_{\frt}$ in which the operators
$\SX_e:=\pi_\frt(\CX_e^{\frt})$ are
{\it positive} self-adjoint. 
By choosing a polarization one may
define 
representations $\pi_\frt$
in terms of multiplication and
finite difference operators on suitable dense subspaces
of the Hilbert
space $\CH_{\frt}\simeq L^2(\BR^{3g-3+n})$.

There exists a family of automorphisms
which describe the relations between the quantized
coordinate functions 
associated to different triangulations \cite{F97,Ka1,CF}. If  
triangulation $\frt_e$ 
is obtained from $\frt$ by changing only the diagonal in the quadrangle
containing $e$, we have 
\begin{equation}\label{q-cluster}
\SX_{e'}^{\frt_e}\,=\,\left\{
\begin{aligned} &\SX_{e'}^{\frt}\,\prod_{a=1}^{|n_{e'e}|}
\big(1+e^{\pi i(2a-1)b^2}          
(\SX_e^{\frt})^{-{\rm sgn}(n_{e'e})}\big)^{-{\rm sgn}(n_{e'e})} \;\;&{\rm if}\;\;e'\neq e\,,\\
&(\SX_e^{\frt})^{-1}\;\;&{\rm if}\;\;{e'= e}\,.
\end{aligned}\right.
\end{equation}
Any two two triangulations $\frt_1$ and $\frt_2$ can be connected by a 
sequence of changes of diagonals in 
quadrilaterals. 
It follows that the quantum theory of $\CM_{\rm flat}^{\CT}(C)$ 
has the structure
of a quantum cluster algebra \cite{FG2}.

It is possible to construct  \cite{Ka1,Ka2} unitary operators
$\ST_{\frt_2\frt_1}:\CH_{\frt_1}\ra\CH_{\frt_2}$ 
that represent the quantum cluster
transformations \rf{q-cluster} in the sense that
\begin{equation}
 \SX_e^{\frt_2}\,=\,
 \ST_{\frt_1\frt_2}^{-1}\cdot \SX_e^{\frt_1}\cdot\ST_{\frt_1\frt_2}\,.
\end{equation}
The operator $\ST_{\frt_2\frt_1}$ describes the change of
representation when passing from the quantum theory
associated to triangulation $\frt_1$ to the one associated
to $\frt_2$. It follows that the 
resulting quantum theory does not depend on the
choice of a triangulation in an essential way.

\subsection{Quantization of the trace functions}\label{algebra}

There is a simple algorithm \cite{F97,FG}
for calculating the trace functions in terms
of the variables $\CX_e^\frt$ leading to Laurent polynomials in the 
variables $\CX_e^\frt$ of the form
\begin{equation}\label{L-X}
\CL_\ga^\frt\,=\,\sum_{\nu\in\BF}C^{\frt}_\ga(\nu)\,
\prod_{e}\,(\CX_e^\frt)^{\frac{1}{2}\nu_e}\,,
\end{equation}
where the summation is taken over a finite set $\BF$ of
vectors $\nu\in\BZ^{3g-3+2n}$ with components $\nu_e$. 

In order to define 
an operator $\SL_\ga^\frt$ associated to a classical 
trace function $\CL_\ga$ it has turned out \cite{CF,CF2,T05}  for some pairs $(\ga,\frt)$ 
to be sufficient to simply replace $(\CX_e^\frt)^{\frac{1}{2}\nu_e}$ in \rf{L-X} by 
$\exp(\sum_{e}\frac{1}{2}\nu_e\log\SX_e^\frt)$. Let us call such pairs $(\ga,\frt)$ simple. In order to define 
$\SL_\ga^{\frt}$ in general \cite{T05} one may use the fact that for all curves $\ga$
there exists a triangulation $\frt'$ such that $(\ga,\frt')$ is simple, allowing us to define 
\begin{equation}
\label{SLtau-inter}
\SL_\ga^{\frt}\,=\,
\ST_{\frt'\frt}^{-1}\cdot \SL_\ga^{\frt'}\cdot\ST_{\frt'\frt}^{}\,.
\end{equation}
The operators $\SL_\ga^\frt$ defined thereby
are positive self-adjoint with spectrum bounded from below by $2$, 
as follows from
the result of \cite{Ka4}. 
Two operators $\SL_{\ga_1}^\frt$ and $\SL_{\ga_2}^\frt$ 
commute if the intersection
of $\ga_1$ and $\ga_2$ is empty.

It turns out that \rf{SLtau-inter} holds in general.
It follows
that we may regard the algebras
generated by the operators $\SL_\ga^\frt$ as different
representations $\pi_\frt$ of an abstract algebra 
$\CA_{b^2}(C)\equiv {\rm Fun}^{\rm alg}_{b^2}(\CM_{\rm flat}^\CT(C))$ 
which does not depend on the
choice of a triangulation, 
$\SL^\frt_\ga\equiv\pi_{\frt}(L_{\ga})$ for $L_\ga\in\CA_{b^2}(C)$.
As in the classical case one may use pants decompositions to identify convenient
sets of generators for $\CA_{b^2}(C)$ to be
\[
\text{set of generators:}\quad
\big\{\, L_i^e,\;i\in\{s,t,u\}\,,e\in\{{\rm edges\;\,of\;\,\Ga}\}\,\big\}\,.
\] 
Important relations are 
\begin{equation}
\begin{aligned}
&\CP^{(a)}_{0,4}(\CL_s^e, \CL_t^e, \CL_u^e;L_1^e,L_2^e,L_3^e,L_4^e)\,=\,0\,,\quad &\text{if}\;\;C_e\simeq C_{0,4}\,,\\
&\CP^{(a)}_{1,1}(\CL_s^e, \CL_t^e, \CL_u^e;L_0^e) \,= \,0\,,\quad &\text{if}\;\;C_e\simeq C_{1,1}\,,
\end{aligned}\quad a=2,3\,.\end{equation}
where the polynomials $\CP^{(a)}_{0,4}$ of non-commutative variables 
are defined as :
\begin{align} \label{CR}
& \CP^{(2)}_{0,4}(\CL_s, \CL_t, \CL_u;L_1,L_2,L_3,L_4):=e^{\pi \textup{i} b^2} \CL_s \CL_t - 
e^{-\pi \textup{i} b^2} \CL_t \CL_s  \\
&\quad 
- (e^{2\pi \textup{i} b^2} - 
e^{-2\pi \textup{i} b^2}) \CL_u - 
(e^{\pi \textup{i} b^2} - e^{-\pi \textup{i} b^2}) (\CL_1\CL_3+\CL_2\CL_4)\,.
\notag\\
& \CP^{(3)}_{0,4}(\CL_s,  \CL_t, \CL_u;L_1,L_2,L_3,L_4) := \CL_1\CL_2\CL_3\CL_4+\CL_1^2+\CL_2^2+\CL_3^2+\CL_4^2 \\  & 
\quad- e^{\pi\textup{i} b^2} \CL_s \CL_t \CL_u
+e^{2\pi \textup{i} b^2} \CL_s^2 + 
e^{-2 \pi \textup{i} b^2} \CL_t^2 + e^{2\pi \textup{i} b^2} \CL_u^2 
-\big(2\cos\pi b^2)^2  \nonumber \\ 
& \quad+ e^{\pi\textup{i} b^2} 
\CL_s (\CL_3\CL_4 + \CL_1\CL_2) + e^{- \pi \textup{i} b^2} 
\CL_t (\CL_2\CL_3 + \CL_1\CL_4) +
e^{\pi \textup{i} b^2} \CL_u (\CL_1\CL_3 + \CL_2\CL_4)\big] \,.
\nonumber
\end{align}
In the case $C_e\simeq C_{1,1}$ we have
\begin{align} \label{CR1,1}
\CP^{(2)}_{1,1}(\CL_s, \CL_t, \CL_u;L_0)&:= e^{\frac{\pi \textup{i}}{2} b^2} \CL_s \CL_t - 
e^{-\frac{\pi \textup{i}}{2} b^2} \CL_t \CL_s  
\,- \,(e^{\pi \textup{i} b^2} - 
e^{-\pi \textup{i} b^2}) \CL_u\,,\\
\CP_{1,1}^{(3)}(\CL_s,  \CL_t, \CL_u;L_0) &:= 
e^{\pi \textup{i} b^2}(\CL_s^2 + e^{-2\pi \textup{i} b^2}\CL_t^2 + 
\CL_u^2) - e^{\frac{\pi \textup{i}}{2} b^2}\CL_s \CL_t \CL_u 
+ L_0-2\cos\pi b^2.
\end{align}
The quadratic relations $\CP^{(2)}_{g,n}=0$ represent the deformation of the 
Poisson bracket \rf{loopPB}, while the cubic\footnote{Relations cubic in $L_s$, $L_t$, $L_u$.} relations $\CP^{(3)}_{g,n}=0$
are deformations of the relations \rf{algrel}. One furthermore finds quantum analogs
of the skein relations \cite{CF,CF2}.

\subsection{Representations associated to pants decompositions}\label{reprs}

The operators $\SL_\ga^\frt$ and $\SL_{\ga'}^\frt$ associated to non-intersecting 
curves $\ga$ and $\ga'$ commute.
It is therefore possible to diagonalise simultaneously the 
quantised trace functions associated to a maximal set $\ga=\{\ga_1,\dots,\ga_h\}$ of 
non-intersecting closed curves
characterising a pants decomposition. 
This can be done by constructing
operators $\SR_{\si|\frt}$ which map
the operators $\SL_{\ga_e}^\frt$ 
associated to the curves  $\ga_e$, $e=1,\dots,h$, to the 
operators of multiplication by  $2\cosh(l_e/2)$, respectively  \cite{T05,TV2}. 
The states in the image $\CH_\si$ of $\SR_{\si|\frt}$
can be represented by functions $\psi(l)$, 
$l=(l_1,\dots,l_h)$
depending on the variables $l_e\in\BR^+$ which parameterise the
eigenvalues of $\SL_{\ga_e}^\frt$.
The operators $\SR_{\si|\frt}$ define a new family of 
representations $\pi_\si$ of $\CA_{b^2}(C)$ via
\begin{equation}
\label{SLtau-sigma}
\pi_\si(L_\ga):=\SR_{\si|\frt}
\cdot \pi_\frt(L_\ga)\cdot(\SR_{\si|\frt})^{-1}\,.
\end{equation}
The representations $\pi_\si$ 
are naturally labelled by pants decompositions
$\si=(\ga,\Ga)$.
The unitary operators $\SR_{\si|\frt}:\CH_\frt\ra\CH_\si$ 
were constructed explicitly
in \cite{T05}.

The operators $\pi_\si(\SL_{\ga})$ were calculated 
explicitly for the
generators of $\CA_{b^2}(C)$ in \cite{TV2}.  When $\si$ corresponds to the pants 
decomposition of $C=C_{0,4}$ 
depicted on the left of Figure \ref{fmove} one finds,
for example, $\SL_s:=2\cosh(\sll/2)$,
\begin{align}
\SL_t:=\,&\frac{1}{ 2(\cosh \sll_s - \cos 2\pi b^2)}
\Big(2\cos\pi b^2(L_2L_3+L_1L_4)+
\SL_s (L_1L_3+L_2L_4)\Big) \\ \label{quantum't Hooft}
& \quad +  \sum_{\ep=\pm 1}
\frac{1}{\sqrt{2\sinh(\sll/2)}}
e^{\ep\sk/2}
\frac{\sqrt{c_{12}(\SL_s)c_{34}(\SL_s)}}{2\sinh(\sll/2)}
e^{\ep\sk/2}
\frac{1}{\sqrt{2\sinh(\sll/2)}}  \,,
\nonumber 
\end{align}
with operators $\sll$ and $\sk$ defined as 
$\sll\,\psi_\si(l)=
l\, \psi_\si(l)$,
$\sk\,\psi_\si(l)=-4\pi{\mathrm i} b^2\,\pa_{l}\psi(l)$,
respectively, while
$c_{ij}(L_s)$ is defined as
$c_{ij}(L_s) =L_s^2+L_i^2+L_j^2+L_sL_iL_j-4$.
$\SL_u$ is given by a similar expression \cite{TV2}. 
The operators $\sll$ and $\sk$ can be identified as quantum counterparts
of the Fenchel-Nielsen coordinates.
In the general case one may 
use pants decompositions to reduce the description of the operators $\pi_\si(\SL_i^e)$,
$i\in\{s,t,u\}$, $e\in\{{\rm edges\;of\;\Ga}\}$ to the cases $C_e\simeq C_{0,4}$ and $C_e\simeq C_{1,1}$.

The operators $\pi_\si(L_\ga)$ are unbounded. The maximal domain of definition
of $\pi_\si(\CA_{b^2}(C))$ defines a natural subspace $\CS_\si\subset \CH_\si$ with topology 
given by the family of semi-norms $\lVert \pi_\si(\CO)\rVert$, $\CO\in\CA_{b^2}(C)$. The 
topological dual $\CD_\si$
of $\CS_\si$ is a space of distributions canonically associated to $(\CA_{b^2}(C),\pi_\si)$
such that $\CS_\si\subset\CH_\si\subset\CD_\si$.

\subsection{Changes of pants decomposition}\label{chpants}

 The passage between the representations $\pi_{\si_1}$ 
and $\pi_{\si_2}$ associated to 
two different pants decompositions can be described by
\begin{equation}
\SU_{\si_2\si_1}:=\SR_{\si_2|\frt}\cdot(\SR_{\si_1|\frt})^{-1}\,.
\end{equation}
The passage between two pants decompositions $\si_1$ 
and $\si_2$ can always be decomposed into elementary "moves" 
called F-, S-, B- and Z- moves 
localized in subsurfaces with $3g-3+n\leq 1$  \cite{MS,BK,FG}. 
We refer to  \cite{BK,FG} for precise descriptions of the full set of generators.
For future reference we have depicted the F- and S- moves in 
Figures \ref{fmove} and \ref{smove}, respectively.  It is useful
to formalize the resulting structure by introducing 
the notion of the Moore-Seiberg groupoid: The path groupoid 
of the two-dimensional CW-complex which has vertices identified with
pants decompositions 
$\si$, edges (``generators'') called F-, S-, B- and Z-moves, 
and faces (``relations'') 
being certain edge-paths localized in subsurfaces with 
$3g-3+n\leq 2$ listed in  \cite{MS,BK,FG}.

The unitary operators $\SU_{\si_2\si_1}$ intertwine the
representation $\pi_{\si_1}$ and $\pi_{\si_2}$,
\begin{equation}\label{inter}
\pi_{\si_2}(\CL_\ga)\cdot \SU_{\si_2\si_1}\,=\,
\SU_{\si_2\si_1}\cdot \pi_{\si_1}(\CL_\ga)\,.
\end{equation}
Explicit representations for the operators 
$\SU_{\si_2\si_1}$ have been calculated in \cite{NT,TV2} 
for pairs $[\si_2,\si_1]$ related by the generators 
of the 
Moore-Seiberg groupoid. The B-move is represented as 
\begin{equation}\label{Bcoeff}
\SB\psi\,=\,B_{l_\2l_\1}^{l_3}\psi_s\,,
\qquad B_{l_\2l_\1}^{l_\3}\,=\,
e^{\pi i(\De_{l_\3}-\De_{l_\2}-\De_{l_\1})}\,,
\end{equation}
where $\De_l=(1+b^2)/4b+(l/4\pi b)^2$, and $\psi$ is a generator for the 
one-dimensional space associated to $C_{0,3}$.
The F-move is represented in terms of an 
integral transformation of the form
\begin{equation}\label{F-trsf}
\psi_s(l_s)\,\equiv\,(\SF\psi_t)(l_s)\,=\,
\int_{\BR^+} dl_t \;\Fus{l_1}{l_2}{l_3}{l_4}{l_s}{l_t}\;
\psi_t(l_t)\,.
\end{equation}
A similar formula exists for the S-move.
The explicit formulae are given in \cite{TV2}.

The 
operators $\SU_{\si_\2\si_1}$ generate 
a projective unitary
representation of the Moore-Seiberg
groupoid, 
\begin{equation}\label{MS-abstr}
\SU_{\si_\3\si_2}\cdot\SU_{\si_\2\si_1}\,=\,\zeta_{\si_3\si_2\si_1}
\SU_{\si_\3\si_2}\,,
\end{equation}
where $\zeta_{\si_3\si_2\si_1}\in\BC$, $|\zeta_{\si_3\si_2\si_1}|=1$.
The explicit formulae for the relations of the 
Moore-Seiberg groupoid in the quantization of $\CM^\CT_{\rm flat}(C)$ 
are listed in \cite{TV2}.
The operators $\SU_{\si_2\si_1}$ allow us to identify the spaces $\CS_\si\subset\CH_\si\subset\CD_\si$
as different representatives of abstract  spaces $\CS(C)\subset\CH(C)\subset\CD(C)$ associated to $C$.

Having a representation of the Moore-Seiberg groupoid 
induces a representation of the mapping class group ${\rm MCG}(C)$. 
Elements $\mu$ of ${\rm MCG}(C)$  can be represented by diffeomorphisms
of the surface $C$ not isotopic to the identity, and therefore map any pants decomposition $\si$ 
to another one denoted $\mu.\si$. Note that the 
Hilbert spaces $\CH_\si$ and $\CH_{\mu.\si}$ are canonically 
isomorphic, depending only on the combinatorics
of the graphs $\si$, but not on their embedding into $C$. We may therefore
define an operator $\SM_\si(\mu):\CH_\si\ra\CH_\si$ as
\begin{equation}\label{MCGdef}
\SM_\si(\mu):=\,\SU_{\mu.\si,\si}\,.
\end{equation}
It is automatic that the operators $\SM(\mu)$ define a projective 
unitary representation of  ${\rm MCG}(C)$ 
on $\CH_\si$.

\subsection{An analog of a modular functor}

The description using representations
associated to pants decompositions has the advantage to make manifest 
that we are dealing with an analog of a modular functor. This means in 
particular that the representations of the mapping class group 
associated to
Riemann surfaces of varying topological type $C_{g,n}$ restrict
to, and are generated by, the representations associated to 
embedded subsurfaces of simple topological type $C_{0,3}$, $C_{0,4}$
and $C_{1,1}$. This property 
can be seen as a locality property that is essential for 
having relations with conformal field theory. However, we are not dealing with
a modular functor in the strict sense axiomatised in the 
mathematical literature (see e.g. \cite{BK2,Tu}): 
The definition is restricted to stable
surfaces ($2g-2+n>0$), and the vector spaces associated to such surfaces
are infinite-dimensional in general. However, the theory
described above still exhibits the most essential features
of a modular functor, it is in many respects as close to a
modular functor as it can be in cases 
where the vector spaces associated to surfaces $C_{g,n}$ are 
infinite-dimensional.

It would interesting to develop a generalised 
notion of modular functor that encompasses the quantum 
Teichm\"uller theory and the many conceivable generalizations.
Some suggestions in this direction were made in \cite{T08}.

\section{Conformal field theory}\label{Vircfbl}

\subsection{Definition of the conformal blocks}

\newcommand{\vir}{{\rm Vir}_{c}}

\newcommand{\CFB}{{\CC\CB}}
\newcommand{\HFB}{{\CH_{\rm\sst CFT}}}
\newcommand{\SFB}{{\CS_{\rm\sst CFT}}}
\newcommand{\DFB}{{\CD_{\rm\sst CFT}}}

The Virasoro algebra ${\rm Vir}_c$ has generators $L_n$,  $n\in\BZ$, satisfying the
relations
\begin{equation}\label{Vir}
[L_n,L_m] = (n-m)L_{n+m}+\frac{c}{12}n(n^2-1)\de_{n+m,0}.
\end{equation}
We will consider irreducible highest weight representations 
$\CV_{\al}$ of ${\rm Vir}_c$ with $c>1$ 
generated from vectors $e_{\al}$ annihilated by all $L_n$, 
$n>0$, having
$L_0$-eigenvalue $\al(Q-\al)$ if $c$ is parameterised as $c=1+6Q^2$. 

Let $C\equiv C_{g,n}$ be a Riemann surface with genus $g$, $n$ marked points 
$P_1,\ldots,P_n$, and choices of local coordinates $t_r$, $r=1,\dots,n$
vanishing at $P_r$, respectively. It will be convenient to assume that the 
local coordinates $t_r$ are part of an atlas of local 
holomorphic coordinates on $C$ with transition functions 
represented by M\"obius-transformations. 
Such an atlas defines a projective structure on $C$.

We associate highest weight representations
$\CV_r\equiv\CV_{\al_r}$, of $\vir$ 
to $P_r$, $r=1,\ldots,n$. 
The conformal blocks
are  linear functionals
$\CF:\otimes_{r=1}^n\CV_r\ra\BC$ 
satisfying the invariance property
\begin{equation}\label{cfblvir}
\CF(\rho_\chi v)\, = 0\,,\quad \forall v\in\otimes_{r=1}^n\CV_r,\quad \forall\chi\in
\FV_{{\rm out}}(C)\,,
\end{equation}
where the notation $\FV_{{\rm out}}(C)$ is used for the Lie algebra of meromorphic
differential operators on $C$ which may have poles only at
$P_1,\ldots,P_n$.   The representation $\rho$ of $\FV_{{\rm out}}(C)$ is defined on 
$\otimes_{r=1}^n\CV_r$ via
\begin{align}\label{Tdef}
\rho_\chi= \sum_{r=1}^n \sum_{k\in\BZ}\chi_k^{(r)}L_k^{(r)}\,,\qquad
L_k^{(r)}:={\rm id}\ot\dots\ot\underset{(\rm r-th)}{L_k^{}} \ot\dots\ot{\rm id}\,,
\end{align}
where the $\chi_k^{(r)}$ are the Laurent coefficients
of $\chi$ at $P_r$,
$\chi(t_r) = \sum_{k\in\BZ} \chi_k^{(r)}\,t_r^{k+1} \,\pa_{t_r} \in
\BC(\!(t_r)\!)\pa_{t_r}$.
We may refer e.g. to \cite{BF} for more details.

The vector space $\CFB(C,\rho)$
of conformal blocks associated to the Riemann surface $C$
is the space of solutions to 
the defining invariance conditions \rf{cfblvir}. 
The space $\CFB(C,\rho)$ infinite-dimensional in general,
being isomorphic to the 
space of {\it formal} power series in $3g-3+n$ variables.

{\it Example.} Let $n=1$. Using the Weierstrass gap theorem
it is straightforward to show that 
the defining condition \rf{cfblvir} allows us to express
the values $\CF(v)$ for any $v\in\CV_0$ in terms
of the collection of complex numbers $
\CF(L_{-h}^{n_{h}}\dots L_{-1}^{n_1}e_{\al_1}\big)$, 
$n_k\in\BZ^{\geq 0}$, $k=1,\dots, h$, where $h:=3g-3+1$.

To any conformal block $\CF$, let us associate the {\it chiral partition function} defined
as the value 
\begin{equation}\label{Z-def}
\CZ(\CF):=\, 
\CF(e)\,,\qquad e:={\textstyle \otimes_{r=1}^n e_{\al_r}}.
\end{equation}

The vacuum representation $\CV_0$ corresponding to
$\al=0$ plays a distinguished role. It can be shown
that the spaces of conformal blocks with and without
insertions of the vacuum representation are canonically
isomorphic, see e.g. \cite{BF} for a proof. Let the surface $C'$ be obtained from $C$
by introducing an additional marked marked point $P_0$.
Let $\rho$ and $\rho'$ be the 
representations of $\FV_{{\rm out}}(C)$ and $\FV_{{\rm out}}(C')$,
defined above
on $\otimes_{r=1}^n\CV_r$ and $\CV_0\ot(\otimes_{r=1}^n\CV_r)$, 
respectively.
To each $\CF'\in\CFB(C',\rho')$
one may then associate a conformal block
$\CF\in\CFB(C,\rho)$ such that 
\begin{equation}\label{vacprop}
\CF'(e_0\ot v) \equiv
\CF(v)\,.
\end{equation}
for all $v\in\otimes_{r=1}^n\CV_r$.
This fact is often
referred to as the ``propagation of vacua''.

\subsection{Deformations of the complex structure of $C$}\label{FSconn}

We shall now discuss the dependence of the spaces of conformal blocks
on the choice of the Riemann surface $C$.
The definition above defines sheaves of conformal blocks over $\CM_{g,n}$,
the moduli space of complex structures on surfaces of genus $g$ and
$n$ punctures.
Let us consider a local patch $\CU\subset \CM_{g,n}$ parameterised by local 
complex analytic
coordinates $q=(q_1,\dots,q_{3g-3+n})$, and represented by 
families $C_q$ of Riemann surfaces with holomorphically varying 
projective structures. 

A basic observation concerning the dependence of 
the space of conformal blocks on the complex structure
is the existence of a canonical connection on the sheaves of conformal blocks
over $\CM_{g,n}$. Let us define
the infinitesimal variations 
\begin{equation}\label{Viract}
\de_{\chi} \CF(v) := \CF(\rho_\chi v)\,,
\end{equation}
with $\rho_\chi$
being defined via \rf{Tdef} for arbitrary
$\chi\in\oplus_{k=1}^n \BC(\!(t_k)\!)\pa_{t_k}$.
The ``Virasoro uniformization theorem'' (see e.g. \cite{BF} for a proof)
implies that the
Teichm\"uller space, 
being the tangent space $T{\CM}_{g,n}$ to the space
of complex structures ${\CM}(C)$ at $C$ 
is isomorphic to the double quotient
\begin{equation}\label{VirUni}
\CT(C) \simeq \FV_{{\rm out}}(C) 
\left\backslash \,\oplus_{k=1}^n \BC(\!(t_k)\!)\pa_{t_k}\,
\right/ \FV_{{\rm in}}(C)\,; 
\end{equation}
$\BC(\!(t_k)\!)$ denotes the space of finite Laurent series, while $\FV_{{\rm in}}(C):=
\oplus_{k=1}^n \BC[[t_k]]\pa_k$, with $\BC[[t_k]]$ 
being the space of finite Taylor series in the variable $t_k$. 
Assuming  temporarily $\al_r=0$, $r=1,\dots,n$, it follows from
\rf{VirUni} that that \rf{Viract}
relates the values $\CF(\rho_\chi e)$ to 
derivatives of the chiral partition functions $\CZ(\CF)$ with respect
to the complex structure moduli of $C$. More general cases for the parameters $\al_r$ 
can be treated similarly. Using the propagation of vacua
one may use \rf{Viract}, \rf{VirUni} to define a differential 
operator $\ST(y)$ on $\CT(C)$ such  that
\begin{equation}\label{Viract2}
\ST(z_0)\CF(v)\,=\,\CF'(L_{-2}e_0 \ot v)\,.
\end{equation}
The defining 
conditions \rf{cfblvir}, \rf{Viract} imply that the conformal 
blocks $\CF$ are fully characterised by the 
collection of all multiple derivatives 
of  $\CZ(\CF)$. 

There are two obstacles  to the integration of  the canonical connection
on  $\CFB(C,\rho)$, in general.
The first problem is that the connection defined by \rf{Viract} is not
flat, but only projectively flat. 
One may, however, trivialize the curvature at least locally,
opening the possibility to integrate
\rf{Viract} at least in open subsets 
$\CU\subset\CM_{g,n}$. We will later 
define  sections  horizontal with respect to the
canonical connection using the gluing 
construction of
conformal blocks.

The other problem is that  $\CFB(C,\rho)$ is simply way too big, 
multiple derivatives defined 
via \rf{Viract} may grow without bound.
However, there exist interesting subspaces of $\CFB(C,\rho)$ on which 
the canonical connection may be integrated.
Let $\CFB^{\rm an}(C,\rho)$ be the subspace of 
$\CFB(C,\rho)$ such that $\CZ(\CF_{C_q})\equiv \CZ(\CF_q)$ can be analytically
continued over all of $\CT(C)$. 
Projective flatness of the canonical connection implies that we
may in this way define a
projective representation of the 
mapping class group on $\CFB^{\rm an}(C,\rho)$.
We will later briefly describe nontrivial evidence for the 
existence of a Hilbert-subspace  $\HFB(C,\rho)$
of  $\CFB^{\rm an}(C,\rho)$
which is closed under this action.  
The projective representation 
of the mapping class group on $\HFB(C,\rho)$ will then 
define an infinite-dimensional unitary projective local 
system $\CW(C)$ over $\CM(C)$. 
This seems to be the best possible
scenario one can hope for when the spaces of 
conformal blocks are infinite-dimensional.

It is known \cite{FS} 
that the projectiveness of the local systems originating
from the canonical connection on spaces of conformal blocks can
be removed by tensoring with the projective line bundle 
$\CE_c=(\la_H)^{\frac{c}{2}}$, where $\la_H$ is the Hodge line bundle.
It follows that 
$\CV(C):=\CW(C)\ot \CE_c$ 
is an ordinary holomorphic vector bundle over $\CM(C)$.
We are next going to describe how to construct {\it global} sections
of $\CV(C)$ by means of the gluing construction.

\subsection{Gluing construction of conformal blocks}\label{glue}

We are now going to  construct large families of
conformal blocks by means of the 
gluing construction. 

\subsubsection{Gluing Riemann surfaces}

Let $C$ be a possibly disconnected Riemann surface, $q\in\BC$ with $|q|<1$, and $D_i(q):=
\{P\in C;|z_i(P)|<|q|^{-\frac{1}{2}}\}$, $i=1,2$ be non-intersecting
discs with local coordinate $z_i(P)$ vanishing at points $P_{0,i}$, for $i=1,2$, respectively.
Let us then define a new Riemann surface $C^\sharp$ by identifying 
the annuli $A_i(q):=\{P\in C;|q|^{\frac{1}{2}}<|z_i(P)|<|q|^{-\frac{1}{2}}\}$
iff the coordinates $z_i(Q_i)$ of points $Q_i\in A_i$ satisfy
$
z_1(Q_1)z_2(Q_2)=q.
$
The gluing parameter $q$ becomes part of 
complex structure moduli
of $C^\sharp$.
By iterating this construction one may build Riemann surfaces $C_{g,n}$ of arbitrary genus $g$ and 
arbitrary number $n$ of punctures from
three-punctured spheres $C_{0,3}$. 

The surfaces $C_{g,n}$ obtained in this way come
with a collection of embedded annuli $A_r(q_r)$, $r=1,\dots,h$, $h:=3g-3+n$.  
As the complex structure on $C_{0,3}\simeq \BP^1\setminus \{0,1,\infty\}$ is unique,
one may use $q=(q_1,\dots,q_{h})$ as local coordinates
for $\CM_{g,n}$ in a multi-disc centered around the boundary component in the
Deligne-Mumford compactification $\overline{\CM}_{g,n}$ of ${\CM}_{g,n}$
represented by the nodal surface obtained in the limit ${\rm Im}(q_r)=0$, $r=1,\dots,h$.
It is possible to cover $\overline{\CM}_{g,n}$  by local charts 
corresponding to the pants decompositions of $C$ \cite{HV}.
In order to get local coordinates for the Teichm\"uller spaces $\CT(C)$
one may parameterise $q_r=e^{2\pi i\tau_r}$.
Different local charts $\CU_\si\subset\CT(C)$ defined by the gluing construction
can be labelled by the pairs $\si=(\ga,\Ga)$ introduced in Section \ref{Dehn}.

Using the coordinates around the punctures of $C_{0,3}$ 
coming from the representation 
as $\BP^1\setminus \{0,1,\infty\}$ in the gluing construction 
one gets an atlas on $C$ 
with transition functions represented by M\"obius-transformations
defining
a projective structure on $C_{g,n}$. By varying the gluing parameters 
$q_r$ one gets local 
holomorphic sections of the affine bundle $\CP(C)$ 
of projective structures over $\CU_\si$.

\subsubsection{Gluing conformal blocks}\label{glueblocks}

Let us first consider Riemann surfaces $C_{2}\sharp C_1$
obtained by gluing two surfaces $C_i$
with $n_i+1$, $i=1,2$ boundary components, respectively.
Let $n=n_1+n_2$, and let $I_1$, $I_2$ be sets such that
$I_1\cup I_2=\{1,\ldots,n\}$. 
Let $\CF_{C_i}\in\CFB(C_i,\rho_i)$, $i=1,2$ be 
conformal blocks with $\rho_i$ acting on 
$\CV^{[n_i]}_i=\CV_\be\otimes(\otimes_{r\in I_i}\CV_r)$
for $i=1,2$, respectively. Let
$\langle\,.,.\,\rangle_{\CV_\be}$ be the $\vir$-invariant bilinear
form on $\CV_\be$, and $\{v_\jmath;\jmath\in\BI_\be\}$, $\{\check{v}_\jmath;\jmath\in\BI_\be\}$
be dual bases for $\CV_\be$ satisfying $\langle v_\jmath,\check{v}_{\jmath'}\rangle_{\CV_\be}=\de_{\jmath,\jmath'}$.
For given $v_i\in\otimes_{r\in I_i}\CV_r$
let $V_i(v_i)$ be the vectors in $\CV_\be$ defined by
\begin{equation}\label{Vprefactor}
V_i(v_i) := \sum_{\jmath\in \BI(\CV_\be)}
\check{v}_{\jmath}\;\CF_{C_i}({v}_\jmath\ot v_i)\,.
\end{equation}
A conformal block associated to
the surface $C_{2}\sharp C_1$ can then be constructed as
\begin{equation}\label{Vfactor}
\CF_{C_{2}\sharp C_1}^\be(v_2\ot v_1) := \big\langle\, V_2(v_2)\,,\,
q^{L_0}V_1(v_1)\big\rangle_{\CV_{\be}}\,.
\end{equation}

An operation representing 
the gluing of two boundary components of a single Riemann surface
can be defined in a very similar way.

\subsubsection{Gluing from pairs of pants}\label{gluingpants}

One can construct any Riemann surface $C$ by gluing pairs of
pants. Different ways of doing this are
labelled by pants decompositions $\si$. The building 
blocks, the conformal blocks associated to $C_{0,3}$,
are uniquely defined
by the invariance property \rf{cfblvir}  up to
the value of $\CF_{C_{0,3}}$ on the product of 
highest weight vectors 
\begin{equation}\label{normcond}
N(\al_3,\al_2,\al_1):=
\CF_{C_{0,3}}(e_{\al_3}\ot e_{\al_2} \ot e_{\al_1})\,.
\end{equation}
Using the
gluing construction recursively leads to the definition of
a family of conformal blocks
$\CF^{\si}_{\be,q}$
depending on the choice of pants decomposition $\si=(\ga,\Ga)$,
the coordinate $q$ for $\CU_{\si}\subset\CT(C)$ defined 
by the gluing construction, and an
assignment $\be:e\mapsto \be_{e}\in 
\BC$ of complex numbers
to the edges $e$ of $\Ga$.  The parameters
$\be_{e}$ determine the Virasoro representations
$\CV_{\be_{e}}$ to be used in the gluing
construction.

The partition functions $\CZ_{\si}(\be,q)$
defined from $\CF^\si_{\be,q}$ via \rf{Z-def} represent
local sections of $\CV(C)$ which are 
horizontal with respect to the canonical connection defined
in Section \ref{FSconn}.

\subsubsection{Change of pants decomposition}\label{changepants}

It turns out that the partition functions 
$\CZ_{\si_1}(\be,q)$ constructed by the gluing
construction in a neighborhood of the asymptotic region of
$\CT(C)$ that is determined by $\si_1$ have an analytic
continuation to
the asymptotic region of $\CT(C)$ determined by a second
pants decomposition $\si_2$. Based on \cite{T01,T03a} 
it was proposed in  \cite{TV2} that
the analytically continued partition functions
$\CZ_{\si_\1}(\be_\1,q)$ are related to 
the functions $\CZ_{\si_\2}(\be_\2,q)$ by linear 
transformations of the form
\begin{equation}\label{CBTrans}
\CZ_{\si_\1}(\be_\1,q) \,=\,E_{\si_\1\si_\2}(q)
\int d\mu(\be_\2)\;W_{\si_1\si_2}(\be_\1,\be_\2)\,
\CZ_{\si_\2}(\be_\2,q)\,.
\end{equation}
The transformations \rf{CBTrans} define the 
infinite-dimensional vector bundle 
$\CV(C)=\CE_c\ot\CW(C)$. The
constant kernels $W_{\si_1\si_2}(\be_\1,\be_\2)$ represent the transition
functions of the projective local 
system $\CW(C)$, while the pre-factors $E_{\si_1\si_2}(q)$
can be identified as transition functions of the 
projective line bundle $\CE_c$.

It is enough to establish \rf{CBTrans} for the cases $C=C_{0,4}$ and $C_{1,1}$
since the Moore-Seiberg groupoid is generated from
the F-, S-, B- and Z-moves.
A partly conjectural\footnote{The main conjecture is the integrability of the representation
of the algebra \cite[equation (201)]{T01}, equivalent to the 
validity of the representation \cite[equation (202)]{T01}.}
argument was proposed in \cite{T01,T03a} suggesting that the F-move can be realised by an 
integral transformation of the form
\begin{equation}\label{fustrsf}
\CZ_{\si_s}(\be_1,q) \,=\,
\int_\BS d\be_2\;\Fus{\al_1}{\al_2}{\al_3}{\al_4}{\be_{1}}{\be_2}\,
\CZ_{\si_t}(\be_2,q)\,;
\end{equation}
where $\BS:=\frac{Q}{2}+i\BR^+$. The relevant pants decompositions 
$\si_s$ and $\si_t$ are depicted on 
the left and right half of Figure \ref{fmove}, respectively. We assume that $\be_1\in\BS$, and that
the parameters
$\al_i\in\BS$, $i=1,2,3,4$ label the representations assigned to the boundary 
components of $C_{0,4}$ according to the labelling indicated in Figure \ref{fmove}.

It was shown  in \cite{HJS} that
\rf{fustrsf} implies the following realisation of the S-move 
\begin{equation}\label{Strsf-L}
\CZ_{\si_s}(\be_1,q) 
\,=\,e^{\pi i\frac{c}{12}(\tau+1/\tau)}
\int_\BS d\be_2\;S_{\be_1\be_2}(\al_0)\,
\CZ_{\si_t}(\be_2,q)\,,
\end{equation} 
The pants decompositions $\si_s$ and $\si_t$ 
for $C=C_{1,1}$ are 
depicted in Figure \ref{smove}.

\section{Comparison with the quantization of the moduli spaces of flat connections}
\label{comparison}

One may now compare the representation of the Moore-Seiberg groupoid 
obtained from the quantisation of $\CM^{\CT}_{\rm flat}(C)$ to the one 
from conformal field theory. It turns out that one finds exact agreement if (i) the representation
parameters are identified 
as 
\begin{equation}\label{parid}\be_e=\frac{Q}{2}+i\
\frac{l_e}{4\pi b}\,,\qquad \al_r=\frac{Q}{2}+i\
\frac{l_r}{4\pi b}\,,\qquad
Q=b+b^{-1}\,,
\end{equation} 
where $r=1,\dots,n$, 
respectively,
and if (ii) a suitable normalisation constant
$N(\al_3,\al_2,\al_1)$ is chosen\footnote{
$N(\al_3,\al_2,\al_1)=(C(Q-\al_3,\al_2,\al_1))^{\frac{1}{2}}$,
where   $C(\al_3,\al_2,\al_1)$ is the function defined in \cite{ZZ}.} in \rf{normcond}.
This implies 
that there are  natural  Hilbert-subspaces 
$\HFB(C,\rho)$  of the spaces of conformal blocks $\CFB(C,\rho)$
on which the mapping class group action 
is unitary. These subspaces have (distributional) bases generated 
by the conformal blocks $\CF^\si_{\be,q}$ constructed by the gluing 
construction with $\be_e\in\BS$ for all edges $e$ of $\si$. The Hilbert spaces 
$\HFB(C,\rho)$ are 
isomorphic as representations of the Moore-Seiberg 
groupoid to the Hilbert spaces
constructed in the quantisation of $\CM^\CT_{\rm flat}(C)$ in 
Section \ref{q-Mflat}.

In the rest of this section we will compare the representations of {\it two}
algebras of operators that arise naturally in the two cases, respectively:
The first is the algebra $\CA_{b^2}(C)$ generated by the quantised trace functions.
This algebra will be realised naturally on spaces of conformal blocks in
terms of the so-called Verlinde loop operators \cite{AGGTV,DGOT}. The second
algebra of operators is the algebra of holomorphic differential operators on the
Teichm\"uller spaces $\CT(C)$. This algebra is naturally realised on the 
conformal blocks via \rf{Viract}. We will briefly discuss, following \cite{TV2},
how a natural realisation is motivated from the point of view of the quantisation
of $\CM_{\rm flat}^\CT(C)$.

\subsection{Verlinde line operators}\label{degpunct}

We shall now define a family of operators $\SL_\ga$ called 
Verlinde line operators  labelled by closed curves $\ga$ on $C$ acting on spaces of conformal 
blocks. It will turn out that the operators
$\SL_\ga$ generate a representation of the algebra $\CA_{b^2}(C)$ on the spaces of 
conformal blocks isomorphic to the one from the quantisation of $\CM_{\rm flat}^\CT(C)$.
To define the operators $\SL_\ga$ we will need a few preparations of interest in their own
right.

\subsubsection{Analytic continuation}

The kernels $W_{\si_1\si_2}(\be_\1,\be_\2)$ representing the transformations
have remarkable analytic properties both with respect to the variables
$\be_2$, $\be_1$, and with respect to the parameters $\al_r$, $r=1,\dots,n$
of the representations assigned to the marked points \cite{TV2}.
An argument has furthermore been put forward 
in \cite{T03a} indicating the absolute convergence of the 
series representing $\CZ_{\si}(\be,q)$. If the normalisation
constant in \rf{normcond} is chosen to be $N(\al_3,\al_2,\al_1)\equiv 1$, 
one may then
show that
the functions $\CZ_{\si}(\be,q)$ are entire
in the variables $\al_r$, and meromorphic
in the variables $\be_e$, having poles only if 
$\be_e\in\BD$, where $\BD:=
-\frac{b}{2}\BZ^{\geq 0}-\frac{1}{2b}\BZ^{\geq 0}$.

This suggests that one may embed the space $\HFB(C,\rho)$ into a larger 
space $\DFB(C,\rho)$ which
contains in particular the conformal blocks constructed
using the gluing construction for generic {\it complex} $\be_e\notin \BD:=
-\frac{b}{2}\BZ^{\geq 0}-\frac{1}{2b}\BZ^{\geq 0}$. We will later characterise the spaces 
$\DFB(C,\rho)$ more precisely. We may note, however, that the analytic 
properties of $W_{\si_1\si_2}(\be_\1,\be_\2)$ and $\CZ_{\si}(\be,q)$
ensure that the relations \rf{CBTrans} can be analytically continued. 
The resulting
relations characterise the realisation of the Moore-Seiberg groupoid 
on the spaces 
$\DFB(C,\rho)$.

\subsubsection{Degenerate punctures}

The representations $\CV_{\al}$ with $\al\in\BD$ are called degenerate
expressing the fact that the vectors in $\CV_{{\al}}$ satisfy additional relations.
Most basic are the cases where $\al=0$, and $\al=-b^{\pm 1}/2$. In the first 
case one has $L_{-1}e_{0}=0$, in the second case 
$(L_{-1}^2+b^{\pm 2}L_{-2})e_{-b^\pm/{2}}=0$.

Let $C'$ be obtained from $C$ by introducing an additional marked 
point $z_0\in C$. 
Analytically continuing conformal blocks with respect to the
parameters $\al_r$, $r=0,\dots,n$ allows one, in particular,
to consider the cases where, for example, $\al_0\in\BD$. 
If $\al_0=0\in\BD$, it turns out that 
$\DFB(C',\rho')\simeq \DFB(C,\rho)$, as required
by the propagation of vacua. In the cases $\al_0=-b^{\pm 1}/2$ it can be
shown that
the partition functions $\CZ(\CF_q')$ for $\CF_q'\in\DFB(C',\rho')$ 
satisfy 
equations 
of the form
\begin{equation}\label{nvec}
\big[\,\pa_{z_0}^2+b^{\pm 2}\ST(z_0)\big]\CZ(\CF_q')\,=\,0\,,
\end{equation}
where $\ST(z_0)$ is a certain first order differential operator
that transforms under changes of local coordinates on $C$ as a quadratic
differential\footnote{Remember that we had fixed a family 
of reference projective
structures in the very beginning.}.
We will refer to these equations as the 
Belavin-Polyakov-Zamolodchikov (BPZ-) equations.
It follows in particular
that $\DFB(C',\rho')\simeq \BC^2\ot\DFB(C,\rho)$, with the  two linearly independent
solutions of \rf{nvec} corresponding to the two elements of a basis for
$\BC^2$.

\subsubsection{Definition of the Verlinde line operators}

Consideration of multiple degenerate punctures reveals some interesting 
phenomena.
If, for example, $C''$ is obtained from $C$ by introducing {\it two} 
additional punctures at $z_0$ and $z_{-1}$ with $\al_0=\al_{-1}=-b/2$ 
one finds a subspace of 
$\DFB(C'',\rho'')\simeq \BC^2\ot\BC^2\ot\DFB(C,\rho)$ naturally isomorphic to $\DFB(C,\rho)$. 
This  is similar (in fact related) to the fact 
that the tensor product 
of two two-dimensional representations of $\fsl_2$ contains a one-dimensional
representation.
This phenomenon  allows us to define 
natural embeddings and projections 
\begin{equation}
\begin{aligned}
\imath:\DFB(C,\rho)\,\hookrightarrow \,\DFB(C'',\rho'')\,,\\
\wp:\DFB(C'',\rho'')\,\ra\,\DFB(C,\rho)\,.
\end{aligned}
\end{equation}
Note furthermore that the mapping class group ${\rm MCG}(C)$ 
contains elements $\mu_\ga$
labelled by closed curves $\ga$ on $C$, 
corresponding to the variation of the position 
of $z_0$ along $\ga$. This allows us to define a natural family of operators
on the spaces $\DFB(C,\rho)$ as
\begin{equation}
\SL_{\ga}:=\,\wp\circ\SM(\mu_\ga)\circ\imath\,,
\end{equation}
where $\SM(\mu)$ is the operator representing $\mu$ on $\DFB(C'',\rho'')$. 
The operators $\SL_\ga$ are called Verlinde line operators. 

Comparing 
the explicit formulae for the Verlinde line operators calculated in 
 \cite{AGGTV,DGOT} with the formulae for the operators $\pi_\si(L_\ga)$
found in \cite{TV2} (see Section \ref{reprs}  above) one finds a precise
match. This means that there is a natural action of the algebra $\CA_{b^2}(C)$ of 
quantised trace functions on spaces of conformal blocks.
This action naturally defines dense subspaces 
$\SFB(C,\rho)\subset\HFB(C,\rho)$ as maximal domains of definition
for $\CA_{b^2}(C)$ such that $\DFB(C,\rho)$ is the dual space of distributions 
forming a so-called Gelfand-triple
$\SFB(C,\rho)\subset\HFB(C,\rho)\subset\DFB(C,\rho)$. The spaces 
$\SFB(C,\rho)$ and $\DFB(C,\rho)$ are isomorphic as $\CA_{b^2}(C)$-modules to the spaces 
$\CS(C)$ and $\CD(C)$ introduced in Section \ref{chpants}, respectively.

\subsection{K\"ahler quantization  of $\CT(C)$}

The relation between conformal field  theory and the quantisation of $\CT(C)$ can
be tightened considerably by considering an alternative quantisation scheme for 
$\CT(C)$ \cite{T,TV2} that we shall now discuss. 
Teichm\"uller theory allows one to equip $\CT(C)$ with natural 
complex and symplectic structures. The natural symplectic form 
$\Omega_{\rm WP}$ on 
$\CT(C)$ coincides
with the restriction of the symplectic form $\Omega_{\rm AB}$ on $\CM_{\rm flat}(C)$
to the Teichm\"uller component $\CM^{\CT}_{\rm flat}(C)$. 
Natural functions on $\CT(C)$ are given by the values of the 
quadratic differential $t(y)\equiv t(y|q,\bar q)$ defined
from the metric of constant negative
curvature $e^{2\vf}dyd\bar{y}$ on $C$ as $t(y)=-(\pa\vf)^2+\pa^2\vf$. 
One may find a basis $\{\vartheta_r;r=1,\dots,h\}$ for the space 
$H^0(C,K^2)$ of holomorphic quadratic differentials on $C$ such that the 
functions $H_r\equiv H_r(q,\bar{q})$ on $\CT(C)$ 
defined via $t(y)=\sum_{r=1}^h \vartheta_r(y)H_r$, 
are canonically conjugate to the complex analytic coordinates $q_r$ on $\CT(C)$
in the sense that $\{H_r,q_s\}=\de_{r,s}$
\cite{TZ87b,TT}.

In the corresponding quantum theory it is natural to realize
the operators $\SH_r$ corresponding to $H_r$ 
as differential operators $b^2{\pa_{q_r}}$,  and
to represent states by holomorphic wave-''functions''
$\Psi^\si(q)$ \cite{TV2}\footnote{More precisely holomorphic 
sections of the projective line 
bundle $\CE_c$. $\Psi^\si(q)$ depends on the choice of a pants decomposition as
the definition of the coordinates $q$ depends on it.}. The operator
corresponding to the quadratic differential $t(y)$ will
be a differential operator $\ST(y)$. 
This operator coincides with
the operator 
defined in \rf{Viract2}. Recall that the space of conformal blocks 
$\CFB^{\rm an}(C,\rho)$ can be identified with
the space of holomorphic 
functions on $\CT(C)$. 
These observations suggest us
to identify the space of states in the
quantum theory of $\CT(C)$
with suitable subspaces of $\CFB^{\rm an}(C,\rho)$.

It is natural to require that the mapping class
group is represented on the wave-''functions'' $\Psi^\si(q)$ 
as deck-transformations
$(\SM(\mu)\Psi^\si)(q)=\Psi^\si(\mu.q)$, where $\mu.q$ is the
image of the point $q$ in $\CT(C)$ under $\mu$. One may then 
show that \cite{T,TV2}
\begin{equation}
\CZ_\si(\be,q)\,=\,\Psi_l^\si(q)\,,
\end{equation}
where $\si=(\ga,\Ga)$, $\ga=(\ga_1,\dots,\ga_h)$, 
$\Psi_l^\si(q)$ is the wave-function of an eigenstate of
the operators $\SL_{\ga_e}$, $e=1,\dots,h$,
and the variables are related via \rf{parid}, respectively.

The observations made in this section indicate that
conformal field theory is nothing but another language for
describing the quantum
theories obtained by quantisation of $\CM^\CT_{\rm flat}(C)$.

\section{Further connections}

The theory outlined above generalises and unifies various themes of 
mathematical research. As an outlook we shall now briefly mention
some of these connections, some of which offer interesting perspectives for
future research.

\subsection{Relation with non-compact quantum groups}

There is an interesting non-compact quantum group called modular double 
of $\CU_q(\fsl(2,\BR))$ \cite{Fa} which is on the algebraic level 
isomorphic to $\CU_q(\fsl_2)$, and has a set of unitary irreducible
representations $\CP_s$, $s\in\BR^+$ characterized by
a remarkable self-duality property: They are simultaneously
representations of $\CU_q(\fsl(2,\BR))$ and $\CU_{\tilde{q}}(\fsl(2,\BR))$
with $\tilde{q}=e^{\pi i/b^2}$ if $q=e^{\pi i b^2}$ \cite{PT1,Fa}. This
family of representations is closed under tensor products \cite{PT2,NT},
and there exists a non-compact quantum group ${\rm SL}_q^+(2,\BR)$
deforming a certain subspace of the space of functions on $SL(2,\BR)$
which has a Plancherel-decomposition into the representations $\CP_s$, 
$s\in\BR^+$, \cite{PT1,Ip}.

There exists strong evidence\footnote{The main open problem is the 
issue pointed out in Section \ref{changepants}.} for an equivalence
of braided tensor categories of Kazhdan-Lusztig type \cite{KL} 
between the 
category of unitary representations of the Virasoro algebra having
simple objects $\CV_{\al}$, $\al\in\BS$, with the 
category having the representations $\CP_s$, $s\in\BR^+$ of
the modular double as simple objects. The kernel representing the 
F-move coincides with the 6j-symbols of the modular double of 
$\CU_q(\fsl(2,\BR))$ \cite{T01}. The complex numbers 
numbers representing the B-move coincide with the eigenvalues 
of the R-matrix of the modular double \cite{BT}.

The results of \cite{TV2} furthermore imply that the braided tensor category of unitary 
representations of the modular double has a natural extension to a {\it 
modular} tensor category.

\subsection{Relations to three-dimensional hyperbolic geometry}

The Teichm\"uller theory has numerous relations to hyperbolic geometry
in three dimensions. Let us consider, for example, the Fenchel-Nielsen 
coordinates $(l_s,\kappa_s)$ and  $(l_t,\kappa_t)$ associated to the
pants decompositions on the left and on the right of Figure 3, 
respectively. It was observed in \cite{NRS} that the generating 
function $\CW(l_s,l_t)$ for the change of Darboux coordinates 
$(l_s,\kappa_s)\leftrightarrow(l_t,\kappa_t)$ for $\CT(C)$, 
defined by the relations 
\begin{equation}\label{CWdef}
\kappa_s=\frac{\pa\CW}{\pa l_s}\,,\qquad
\kappa_t=-\frac{\pa\CW}{\pa l_t}\,,
\end{equation}
coincides with the volume ${\rm Vol}_T(l)$ of the hyperbolic tetrahedron
with edge lengths $l=(l_1,l_2,l_3,l_4,l_s,l_t)$, with 
$l_i$, $i=1,2,3,4$ being the hyperbolic lengths of the 
boundaries of $C_{0,4}$.

It is therefore not unexpected
to find relations to hyperbolic geometry encoded within 
quantum Teichm\"uller theory. Considering the limit $b\ra 0$ of the kernel 
$F(l_s,l_t):=\Fus{l_1}{l_2}{l_3}{l_4}{l_s}{l_t}$ appearing in \rf{F-trsf}
one may show that $\lim_{b\ra 0}b^2 \log F(l_s,l_t)$ is equal to  
the volume ${\rm Vol}_T(l)$ of the hyperbolic tetrahedron considered above.
This follows from the fact 
that \rf{inter} reduces to \rf{CWdef} in the limit $b\ra 0$.
A closely related result was found in \cite{TV2} by direct 
calculation.

Braided tensor categories of representations of compact quantum groups 
can be used to construct invariants of three-manifolds \cite{Tu,BK2}. 
It should be 
interesting to investigate similar constructions using the modular
tensor category associated to the modular double. It seems quite possible
the the resulting invariants are related to the invariants
constructed in \cite{Hik,DGLZ,AK}. If so,
one would get an interesting perspective on the variants of the volume 
conjecture formulated in \cite{Hik,DGLZ,AK}: It could be 
a consequence of the relations between quantum Teichm\"uller theory
and hyperbolic geometry pointed out above, which are natural 
consequences of known relations between  Teichm\"uller theory
and three-dimensional hyperbolic geometry.

\subsection{Relations with integrable models}

There are several connections between the 
mathematics reviewed in this article and the theory of integrable models
We will here describe some
connections to the theory of isomonodromic deformations of certain ordinary
differential equations, for $g=0$ 
closely related to the equations studied by Painlev\'e, 
Schlesinger and Garnier. 
Further connections  are 
described in \cite{BT2,T10}.

\subsubsection{Relations with isomonodromic deformations I}

The limit $b\ra 0$ of the BPZ-equations \rf{nvec} 
is related to isomonodromic deformations \cite{T10}.

Let us consider the case of a Riemann surface $\hat{C}\equiv
C_{g,n+d+1}$ with $n+d+1$
marked points $z_1,\dots,z_n$, $u_1,\dots,u_{d}$ and $y$. For convenience
let us assume that  $u_1,\dots,u_{d}$ and $y$ lie in a single chart of the 
surface $C$ obtained from $\hat C$ by filling $u_1,\dots,u_{d}$ and $y$.
 The resulting loss of generality will not be very essential. 
We associate representations with generic value of the parameter $\al_r$ to $z_r$ for $r=1,\dots,n$,
degenerate representations with parameter $-1/2b$ to the points $u_1,\dots,u_d$, and
a degenerate representation with parameter $-b/2$ to the point $y$. The partition functions 
$\CZ(q)\equiv\CZ(\hat\CF_q)$, $\hat\CF_q\equiv\CF_{\hat{C}_q}$ will then satisfy a system of $d+1$ partial 
differential equations of the form,
\begin{subequations}\label{BPZ}
\begin{align}\label{nvec+}
&\big[\,b^{+2}\pa_{u_k}^2+\ST_k(u_k)\big]\CZ(\CF_q)\,=\,0\,,\quad k=1,\dots,d\,,\\
&\big[\,b^{-2}\pa_{y}^2+\ST_0(y)\big]\CZ(\CF_q)\,=\,0\,.
\label{nvec-}\end{align}
\end{subequations}
In the limit $b\ra 0$ one may solve this system of partial differential 
equation with an ansatz of the form  $\CZ(q)=\exp({\frac{1}{b^2}\CW(q')})\psi(y)(1+\CO(b^2))$,
where $\CW(q')$ does not depend on $y$.
Equation \rf{nvec-} implies that $\psi(y)$ satisfies $(\pa_y^2+t(y))\psi(y)=0$,
where $t(y)=b^2\ST_0(y)\CW(q')$. Equations \rf{BPZ} imply that $v_k:=\pa_{u_k}\CW$
satisfy \begin{equation}\label{clnvec}
v_k^2+t_{k,2}^{}=t_{k,1}^2+t_{k,2}^{}=0\,,\qquad k=1,\dots,d\,,
\end{equation}
with $t_{k,l}$ defined from $t(y)=\sum_{l=0}^{\infty}t_{k,l}(y-u_k)^{l-2}$. It follows that $\pa_y^2+t(y)$
has $d$ apparent singularities at $y=u_k$. Let $\vartheta_k(y)(dy)^2$ be a basis for $H^0(C,K^2)$,
and let us define $H_k$ via $t(y)=\sum_{k=1}^dH_k\vartheta_k(y)$.
In the case $d=3g-3+n$ one has enough equations \rf{clnvec}
to determine the $H_k\equiv H_k(u,v)$  as functions 
of $u=(u_1,\dots,u_d)$ and $v=(v_1,\dots,v_d)$.
 
It is automatic that the monodromy of 
$\pa_y^2+t(y)$ will be unchanged under variations of the complex 
structure of $C$, which is equivalent to  \cite{Ok,Iw}
\begin{equation}
\frac{\pa u_k}{\pa{q_r}}\,=\,\frac{\pa{H_{r}}}{\pa v_k}\,,
\qquad
\frac{\pa v_k}{\pa{q_r}}\,=\,-\frac{\pa{H_{r}}}{\pa u_k}\,,
\end{equation}
$\{\frac{\pa}{\pa q_k};k=1,\dots,d\}$ being the basis for $T\CT(C)$ dual to the basis $\{\vartheta_k,k=1,\dots,d\}$ for 
$H^0(C,K^2)\simeq T^*\CT(C)$. It follows that the system of BPZ-equations \rf{BPZ}
describes a quantisation of the isomonodromic deformation problem \cite{T10}. 

\subsubsection{Relations with isomonodromic deformations II}

A somewhat unexpected relation between conformal blocks
and the isomonodromic deformation problem arises in the limit $c\ra 1$.
A precise relation between the tau-function for Painlev\'e VI and
Virasoro conformal blocks was proposed in \cite{GIL}. A proof of this
relation, together with its generalization to the tau-functions of the 
Schlesinger system was given in \cite{ILT}. The relations 
established in \cite{ILT} are
\begin{equation}\label{taufromZ}
\tau(\la,\kappa;q)\,=\,\sum_{m\in\BZ^N}e^{i\kappa\cdot m}
\CZ_\si(\la+m,q)\,,
\end{equation}
where $N=n-3$, $\CZ_\si(\be,q)$ are the chiral partition functions 
associated to the conformal blocks defined using the gluing construction 
in the case $C=C_{0,n}$, and
$\tau(\la,\kappa;q)$ is the isomonodromic tau-function, 
defined by $H_r=-\pa_{q_r}\tau(\la,\kappa;q)$,
here considered as a function of the monodromy data parameterised in terms 
of Darboux coordinates $(\la,\ka)$ for $\homsl$ closely related to the 
coordinates used in \cite{NRS}.

In order to prove \rf{taufromZ}, the authors of \cite{ILT} consider 
partition functions $\CZ(\CF_q'')$ of conformal blocks $\CF''_q\in
\DFB(C'',\rho'')$ with two additional degenerate
punctures as in Section \ref{degpunct}.
Recall that one gets an action of $\pi_1(C)$ on $\DFB(C'',\rho'')$
from monodromies of one of the degenerate punctures. 
The isomorphism $\DFB(C'',\rho'')\simeq \BC^4\ot\DFB(C,\rho)$
allows us to represent the action of $\pi_1(C)$ on
$\DFB(C'',\rho'')$ in terms of matrices having elements which
are difference operators acting on $\DFB(C,\rho)$. The remarkable 
fact observed in \cite{ILT} is that the appearing difference
operators can be diagonalised simultaneously by a
generalised Fourier-transformation similar to \rf{taufromZ}
(provided that $g=0$ and $c=1$). This observation yields
in particular a new and more effectively computable way to 
solve the classical Riemann-Hilbert problem \cite{ILT}.

\section{Outlook: Harmonic analysis on ${\rm Diff}(S^1)$ ?}

Let $\Pi_j$ be a unitary irreducible representation $\Pi_j:G\ra {\rm End}(\CV_j)$
of a finite-dimensional Lie group on a Hilbert space $\CV_j$ with scalar product 
$( .,.):\CV_j\ot\CV_j\ra\BC$, $j$ being a label for elements in the set
of irreducible unitary representations of $G$. 
Matrix elements 
such as
$({v}_2,\Pi_j(g)v_{1})$, $v_i\in\CV_j$ for $i=1,2$,
play a fundamental role
in the harmonic analysis of the Lie group $G$. They allow us to 
realise the abstract Plancherel
decomposition
$L^2(G)\simeq\int^{\oplus}_{\BU} d\mu(j)\,\CV_j\otimes\CV_j^\dagger$ 
as a generalised Fourier-transformation 
\begin{equation}
f(g)=\int_{\BU} d\mu(j)\;\sum_{\imath,\imath'\in \BI_j}
({v}_\imath,\Pi_j(g)v_{\imath'})\,\tilde{f}_{\imath\imath'}(j)\,,
\end{equation}
with $\{v_\imath;\imath\in\BI_j\}$ being an orthonormal basis for $\CV_j$. 
If the representations
$\CV_j$ contain unique vectors $v_{2}^j$, $v_{1}^j$ invariant
under subgroups $H_2$ and $H_1$, respectively, 
one may similarly represent functions
on the double quotients $H_2\backslash G/H_1$,  as
\begin{equation}\label{Fourier}
f(g)=\int_{\BU} d\mu(j)\;
({v}_2^j\,,\,\Pi_j(g)v_{1}^j)\,\tilde{f}(j)\,.
\end{equation}
The functions $\CY(j,g):=({v}_2^j\,,\,\Pi_j(g)v_{1}^j)$ 
are called spherical or Whittaker functions depending on 
the type of subgroups $H_2$ and $H_1$ under 
consideration. Equation \rf{Fourier} expressed the completeness of the 
functions $\CY(j,g)$ within $L^2(H_2\backslash G/H_1)$.

Turning back to conformal field theory let us consider the 
conformal blocks  
constructed by the gluing construction as described in Section 
\ref{glueblocks}. The partition function $\CZ(\be,q)$ 
can be represented as a matrix element in the form 
$\CZ(\be,q)=\langle V_{2},q^{L_0}V_{1}\rangle$.
We could consider, more generally 
\begin{equation}\label{matel}
\CZ(\be,g)\,=\,\langle V_{2}\,,\,\Pi_\be(g)\,V_{1}\rangle\,,
\end{equation}
where $g\in{\rm Diff}^+(S^1)$, and $\Pi_\be$ is the 
projective unitary representation of ${\rm Diff}^+(S^1)$
related to the representation $\CV_{\be}$ of the Virasoro 
algebra by exponentiation,
$\Pi_\be(e^f)=e^{i\pi_\be(T[f])}$, for $f(\si)\pa_\si=\sum_{n\in\BZ}f_ne^{in\si}\pa_\si$ 
being a real smooth vector field on $S^1$, $T[f]=\sum_{n\in\BZ}f_nL_{n}$. 
Equation
\rf{matel} will define a function on ${\rm Diff}^+(S^1)$ that has  an analytic continuation 
to the natural complexfication of ${\rm Diff}^+(S^1)$, the semi-group of annuli $\An$ 
defined in \cite{Se}.
One should note, however, that the states $V_{2}$ and $V_1$ will be 
annihilated by large sub-semigroups ${\An}_{2}$ and ${\An}_{1}$
of ${\An}$, obtained by exponentiation of the Lie-subalgebras of the Virasoro algebra
generated by vector fields on $S^1$ that extend holomorphically to $(C_i\setminus D_i)\cup A_i$, for $i=1,2$, 
respectively. This means that $\CZ(\be,g)$ 
will be a function 
on the double coset $\An_2\backslash\An/\An_1$ which 
can be identified with an open  subset of
the Teichm\"uller space $\CT(C)$. 

This suggests to view the functions $\CZ(\be,g)$ as analogs of spherical
or Whittaker functions. By taking certain collision limits where
the punctures of $C_{0,4}$ collide in pairs one may even construct
honest Whittaker vectors of the Virasoro algebra from the states
$V_i$, $i=1,2$ \cite{GT}, making the analogy even more
close. From this point of view it is intriguing to compare formula 
\rf{fustrsf} with \rf{Fourier}. It is  tempting to view formula
\rf{fustrsf} as an expression of the possible 
completeness of the functions within a - yet to be defined - space
of ``square-integrable'' functions on $\CT(C)$, which in turn  is 
related to a certain coset of the semigroup $\An$ according to the
discussion above. 

These remarks suggest that the relations between conformal field theory and
the quantisation of the moduli spaces of flat ${\rm PSL}(2,\BR)$-connections observed in 
Section \ref{comparison}
should ultimately be understood as results of "quantisation commutes with reduction"-type.
Quantisation of (a space containing) $T^*G$, $G={\rm Diff}^+(S^1)$  should produce an infinite-dimensional
picture close to conformal field theory. The reduction to the finite-dimensional quantum theory
of the Teichm\"uller spaces is a consequence of the invariances of the vectors $V_i$, $i=1,2$.

\end{document}